\documentclass[12pt,prd,aps,floatfix]{revtex4}

\usepackage{graphicx}
\usepackage{amsmath}
\usepackage{amssymb}
\usepackage{longtable}

\newcommand{\be}{\begin{equation}}
\newcommand{\ee}{\end{equation}}
\newcommand{\bea}{\begin{eqnarray}}
\newcommand{\eea}{\end{eqnarray}}
\newcommand{\bi}{\begin{itemize}}
\newcommand{\ei}{\end{itemize}}
\newcommand{\ben}{\begin{enumerate}}
\newcommand{\een}{\end{enumerate}}
\newcommand{\bt}{\begin{tabbing}}
\newcommand{\et}{\end{tabbing}}
\newcommand{\nn}{\nonumber}

\newcommand{\sgn}{{\rm sgn}}
\newcommand{\calO}{{\mathcal O}}
\newcommand{\Hw}{H_{\rm W}}

\newcommand{\pp}{{p^\prime}}
\newcommand{\bfp}{{\bf p}}

\newcommand{\bfx}{{\bf x}}
\newcommand{\bfr}{{\bf r}}
\newcommand{\bfz}{{\bf 0}}
\newcommand{\bfpp}{{{\bf p}^\prime}}
\newcommand{\bfxp}{{{\bf x}^\prime}}
\newcommand{\bfxpp}{{{\bf x}^{\prime\prime}}}
\newcommand{\dt}{{\Delta x_4}}
\newcommand{\dtp}{{\Delta x_4^\prime}}
\newcommand{\qsqmax}{t_{\rm max}}

\newcommand{\fp}{f_+}
\newcommand{\fplo}{f_{+,0}}
\newcommand{\fpnlo}{f_{+,2}}
\newcommand{\fpnloL}{f_{+,2,L}}
\newcommand{\fpnloB}{f_{+,2,B}}
\newcommand{\fpnnlo}{f_{+,4}}
\newcommand{\fpnnloL}{f_{+,4,L}}
\newcommand{\fpnnloC}{f_{+,4,C}}
\newcommand{\fpnnloB}{f_{+,4,B}}
\newcommand{\fpnnnlo}{f_{+,6}}
\newcommand{\fm}{f_-}
\newcommand{\fmlo}{f_{-,0}}

\newcommand{\fmnnloC}{f_{-,4,C}}
\newcommand{\fmnnloL}{f_{-,4,L}}
\newcommand{\fmnnloB}{f_{-,4,B}}

\newcommand{\fz}{f_0}
\newcommand{\fzlo}{f_{0,0}}

\newcommand{\fzt}{\tilde{f}_0}
\newcommand{\fztlo}{\tilde{f}_{0,0}}
\newcommand{\fztnlo}{\tilde{f}_{0,2}}
\newcommand{\fztnloL}{\tilde{f}_{0,2,L}}
\newcommand{\fztnloB}{\tilde{f}_{0,2,B}}
\newcommand{\fztnnlo}{\tilde{f}_{0,4}}
\newcommand{\fztnnloC}{\tilde{f}_{0,4,C}}
\newcommand{\fztnnloL}{\tilde{f}_{0,4,L}}
\newcommand{\fztnnloB}{\tilde{f}_{0,4,B}}
\newcommand{\fztnnnlo}{\tilde{f}_{0,6}}

\newcommand{\fpz}{f_{\{+,0\}}}
\newcommand{\lambdapp}{\lambda_{+}^\prime}
\newcommand{\lambdazp}{\lambda_{0}^\prime}
\newcommand{\lambdapzp}{\lambda_{\{+,0\}}^\prime}
\newcommand{\dfdtp}{\frac{df_+(t)}{dt}}
\newcommand{\dfdtpnlo}{\frac{df_{+,2}(t)}{dt}}
\newcommand{\dfdtpnloL}{\frac{df_{+,2,L}(t)}{dt}}
\newcommand{\dfdtpnloB}{\frac{df_{+,2,B}(t)}{dt}}
\newcommand{\dfdtpnnlo}{\frac{df_{+,4}(t)}{dt}}
\newcommand{\dfdtpnnloC}{\frac{df_{+,4,C}(t)}{dt}}
\newcommand{\dfdtpnnloL}{\frac{df_{+,4,L}(t)}{dt}}
\newcommand{\dfdtpnnloB}{\frac{df_{+,4,B}(t)}{dt}}
\newcommand{\dfdtpnnnlo}{\frac{df_{+,6}(t)}{dt}}
\newcommand{\dfdtz}{\frac{df_0(t)}{dt}}
\newcommand{\dfdtzt}{\frac{d\tilde{f}_0(t)}{dt}}

\newcommand{\pkpff}{F_{V}^{\{\pi^+,K^+\}}}
\newcommand{\pffnnloC}{F_{V,4,C}^{\pi^+}}
\newcommand{\kpffnnloC}{F_{V,4,C}^{K^+}}
\newcommand{\knffnnloC}{F_{V,4,C}^{K^0}}

\newcommand{\Cppt}{{c_{\pi^+,\pi t}^r}}
\newcommand{\Cpkt}{{c_{\pi^+,K t}^r}}
\newcommand{\Ctt}{{c_{t^2}^r}}
\newcommand{\Ckpt}{{c_{K^+,\pi t}^r}}
\newcommand{\Ckkt}{c_{K^+,K t}^r}
\newcommand{\Ckn}{{c_{K^0}^r}}
\newcommand{\Cplspk}{{c_{+,\pi K}^r}}
\newcommand{\Cplspkeff}{{c_{+,\pi K,\rm eff}^r}}
\newcommand{\Cplspt}{{c_{+,\pi t}^r}}
\newcommand{\Cplskt}{{c_{+,K t}^r}}

\newcommand{\Dpls}{{d_+}}
\newcommand{\Dz}{{d_0}}

\begin{document}

\vspace*{-10mm}
\begin{flushright}
\normalsize
 KEK-CP-357        \\
 OU-HET-928        \\
\end{flushright}

\title{
Chiral behavior of $K\!\to\!\pi l \nu$ decay form factors 
in lattice QCD with exact chiral symmetry
}

\author{
   S.~Aoki$^{a}$, 
   G.~Cossu$^{b}$, 
   X.~Feng$^{c,d}$, 
   H.~Fukaya$^{e}$, 
   S.~Hashimoto$^{f,g}$, 
   T.~Kaneko$^{f,g}$, 
   J.~Noaki$^{f}$ and 
   T.~Onogi$^{e}$
   (JLQCD Collaboration)
}

\affiliation{
   $^a$Yukawa Institute for Theoretical Physics,
   Kyoto University, 
   Kyoto 606-8502, Japan
   \\
   $^b$School of Physics and Astronomy, 
   The University of Edinburgh, 
   Edinburgh EH9 3JZ, United Kingdom
   \\
   $^c$Physics Department, Columbia University, 
   New York, NY 10027, USA
   \\
   $^d$School of Physics, Peking University, Beijing 100871, China
   \\
   $^e$Department of Physics, Osaka University, 
   Osaka 560-0043, Japan
   \\
   $^f$High Energy Accelerator Research Organization (KEK),
   Ibaraki 305-0801, Japan 
   \\
   $^g$School of High Energy Accelerator Science,
   SOKENDAI (The Graduate University for Advanced Studies),
   Ibaraki 305-0801, Japan
}

\date{\today}

\begin{abstract}

We calculate the form factors of 
the $K\!\to\!\pi l \nu$ semileptonic decays 
in three-flavor lattice QCD,
and study their chiral behavior 
as a function of the momentum transfer and the Nambu-Goldstone boson masses.
Chiral symmetry is exactly preserved by using the overlap quark action,
which enables us to directly compare the lattice data 
with chiral perturbation theory (ChPT).
We generate gauge ensembles at a lattice spacing of 0.11~fm
with four pion masses covering 290\,--\,540~MeV
and a strange quark mass $m_s$ close to its physical value.
By using the all-to-all quark propagator,
we calculate the vector and scalar form factors with high precision.
Their dependence on $m_s$ and the momentum transfer
is studied by using the reweighting technique and 
the twisted boundary conditions for the quark fields.
We compare the results for the semileptonic form factors
with ChPT at next-to-next-to leading order in detail.
While many low-energy constants appear at this order,
we make use of our data of the light meson electromagnetic form factors
in order to control the  chiral extrapolation.
We determine the normalization of the form factors 
as $f_+(0) \!=\! 0.9636(36)\left(^{+57}_{-35}\right)$,
and observe reasonable agreement of their shape with experiment.

\end{abstract}

\pacs{}

\maketitle


\clearpage 


\section{Introduction}


The kaon semileptonic decays $K\!\to\!\pi l\nu$ provide
a precise determination of the Cabibbo-Kobayashi-Maskawa (CKM) matrix element 
$|V_{us}|$. The decay rate is given as 
\bea
   \Gamma(K\!\to\!\pi l\nu)
   & = &
   \frac{G_F^2 M_K^5}{192\pi^3}
   C^2 S_{\rm EW}
   \left(1+\delta_{\rm EM}+\delta_{SU(2)}\right)^2
   I
   |V_{us}|^2
   \fp(0)^2,
   \label{intro:decay_rate}
\eea   
where $G_F$ is the Fermi constant, 
the Clebsch-Gordan coefficient $C$ is 1 ($1/\sqrt{2}$) 
for the $K^0$ ($K^{\pm}$) decay,
and $I$ represents the phase space integral.
We denote the short-distance electroweak, 
long-distance electromagnetic (EM),
and isospin-breaking corrections by 
$S_{\rm EW}$, $\delta_{\rm EM}$ and $\delta_{SU(2)}$,
respectively.


The vector form factor $\fp(t)$ is defined from relevant hadronic matrix element
\bea
   \langle \pi(\pp) | V_\mu | K(p) \rangle
   & = &
   (p+\pp)_\mu f_+(t) + (p-\pp)_\mu f_-(t),
   \label{eqn:intro:ff}
\eea
where $t\!=\!(\pp-p)^2$ is the momentum transfer.  
Instead of $f_-(t)$, the scalar form factor
\bea
   f_0(t)
   & = &
   f_+(t) + \frac{t}{M_K^2-M_\pi^2} f_-(t)
   \label{eqn:intro:f0}
\eea
has been widely used in phenomenological analyses of 
the semileptonic decays.
By definition, 
the normalizations of the vector and scalar form factors
coincide with each other at the zero momentum transfer $t\!=\!0$,
\bea
  f_+(0) 
  =
  f_0(0).
  \label{eqn:intro:norm}
\eea
The Ademollo-Gatto theorem~\cite{Behrends-Sirlin,Ademollo-Gatto}
states that SU(3) breaking effects are suppressed 
as $f_+(0)\!=\!1+O((m_l-m_s)^2)$, where 
$m_l\!=\!(m_u+m_d)/2$ and 
$m_{\{u,d,s\}}$ represent the masses of up, down and strange quarks.
The normalization $\fp(0)$ has been therefore used 
as an important input to determine $|V_{us}|$
through the decay rate~(\ref{intro:decay_rate}).


Since the form factors describe 
effects due to the strong interaction at low energy,
a precise calculation of $\fp(0)$
needs a non-perturbative method to study QCD.
A target accuracy is $\lesssim 1$\,\%,
because other experimental and theoretical inputs have been determined 
with a similar or even better accuracy~\cite{CKM16:Moulson,CKM16:WG1}.
Lattice QCD is the only known method
to calculate the form factors 
with controlled and systematically-improvable accuracy.


The phase space integral $I$ encodes 
information about the shape of the form factors, 
namely their $t$ dependence.
The current determination of $|V_{us}|$ employs 
a precise estimate of $I$ obtained from experimental data.
A lattice study of the form factor shape and comparison with experiments
can examine the reliability of the numerical lattice determination of
the normalization $\fp(0)$.
We note that new physics can modify not only the normalization 
but also the shape,
which may therefore provide a different probe of new physics
if both theoretical and experimental data become sufficiently 
accurate in the future.


Lattice calculation of $\fp(0)$ has become mature~\cite{FLAG3,CKM16:Simula,CKM16:WG1}
by realistic simulations at reasonably small pion masses and lattice spacings.
The so-called twisted boundary condition~\cite{TBC} enables us to 
simulate near the reference point $t\!\simeq\!0$.
Although recent lattice studies~\cite{Kl3:Nf3:FNAL+MILC,Kl3:Nf3:RBC/UKQCD,Kl3:Nf4:FNAL+MILC,Kl3:Nf4:ETM} have achieved sub-\% accuracy,
more independent calculations are welcome to 
establish the lattice estimate with such a high precision.
A detailed study of the form factor shape 
based on the next-to-leading order (NLO) 
chiral perturbation theory (ChPT)~\cite{ChPT:SU2:NLO,ChPT:SU3:NLO,KFF:ChPT:SU3:NLO}
and model independent parametrization~\cite{z-param,shape:disp:vector,shape:disp:scalar} is also available~\cite{Kl3:Nf4:ETM}.


A recent trend of precision calculations 
is to directly simulate physical quark masses at $t\!=\!0$.
In this study,
we take a different approach based on our large-scale simulations
with exact chiral symmetry~\cite{JLQCD:overlap:summary}.
By exploiting exact symmetry,
we directly compare our lattice data 
with next-to-next-to-leading order (NNLO) ChPT~\cite{KFF:weak:ChPT:SU3:NNLO:PS,KFF:weak:ChPT:SU3:NNLO:BT},
and determine the normalization $\fp(0)$,
relevant low-energy constants (LECs) in the ChPT Lagrangian,
and study the form factor shape.
We note that chiral symmetry is explicitly violated with 
conventional lattice actions.
The explicit violation makes the direct comparison 
between lattice QCD and NNLO ChPT difficult
because of modified functional form of the ChPT formulae
and additional unknown LECs.


For a rigorous comparison with ChPT,
we calculate the form factors with high precision by using 
the so-called all-to-all quark propagator~\cite{A2A:SESAM,A2A:TrinLat}.
The strange quark mass dependence and 
the form factor shape near the reference point $t\!=\!0$ 
are studied by employing the reweighting technique~\cite{reweight:1,reweight:2}
and the twisted boundary conditions, respectively.
While many unknown LECs appear at NNLO,
we control the chiral extrapolation by making use of our lattice data
of the light meson EM form factors 
obtained in Ref.~\cite{EMFF:JLQCD:Nf3:RG+Ovr}.
We also employ a linear combination of $\fz$ and the decay constant ratio
$F_K/F_\pi$~\cite{KFF:weak:ChPT:SU3:NNLO:BT},
which has a reduced number of the LECs.
Our preliminary analyses have been reported in 
Refs.~\cite{Lat10:JLQCD:Kaneko,Lat11:JLQCD:Kaneko,Lat12:JLQCD:Kaneko,Lat15:JLQCD:Kaneko}.


This paper is organized as follows.
In Sec.~\ref{sec:simulation},
we introduce our method to generate our gauge ensembles
and to calculate relevant meson correlators. 
The kaon semileptonic form factors are extracted 
at our simulation points in Sec.~\ref{sec:ff}.
Section~\ref{sec:chiral_fit} details 
comparison with NNLO ChPT to study the chiral behavior 
of the form factors,
and summarizes the numerical results for 
the normalization, shape and relevant LECs.
We summarize our conclusions in Sec.~\ref{sec:conclusion}.


\section{Simulation method}
\label{sec:simulation}

\subsection{Configuration generation}


We simulate $N_f\!=\!2+1$ QCD using
the overlap quark action \cite{Overlap:NN,Overlap:N}
defined by the Dirac operator
\bea
   D(m_q) 
   & = &
   \left( 1 - \frac{m_q}{2m_0} \right) D(0) + m_q,
   \label{eqn:sim:conf_gen:overlap}
\eea
where $m_q$ represents the quark mass, and 
\bea
   D(0) 
   & = & 
   m_0 \left( 1 + \gamma_5 \, \sgn \left[ \Hw (-m_0) \right] \right)
   \label{eqn:sim:conf_gen:overlap:m0}
\eea   
is the overlap-Dirac operator in the massless limit.
The parameter $m_0$ for the Hermitian Wilson-Dirac operator $H_W$
is set to $m_0\!=\!1.6$ 
from our study of the locality of $D$~\cite{Prod_Run:JLQCD:Nf2:RG+Ovr}.
Numerical simulations are remarkably accelerated 
by modifying the Iwasaki gauge action~\cite{Iwasaki} 
with an auxiliary Boltzmann weight~\cite{exW:Vranas,exW+extmW:JLQCD}
\bea
   \Delta_{\rm W} 
   & = & 
   \frac{\det[\Hw(-m_0)^2]}{\det[\Hw(-m_0)^2+\mu^2]}
   \hspace{5mm}
   (\mu=0.2)
   \label{eqn:sim:conf_gen:det}
\eea
and by simulating the trivial topological sector.
The effect of the fixed {\it global} topology 
can be considered as a finite volume effect
suppressed by the inverse lattice volume~\cite{fixed_Q:AFHO}.
In fact, the effect turns out to be small 
in our previous study of the pion EM form factor
on similar and even smaller lattice volumes~\cite{PFF:JLQCD:Nf2:RG+Ovr,EMFF:JLQCD:Nf3:RG+Ovr}.
We also note that 
local topological excitations are active in our gauge ensembles.
Indeed, 
the topological susceptibility calculated in our simulations
is nicely consistent with 
the prediction of ChPT~\cite{chi_t:JLQCD}.


The lattice spacing $a$ determined from the $\Omega$ baryon mass
is $0.112(1)$\,fm
with our choice of the gauge coupling $\beta\!=6/g^2\!=\!2.30$.
We simulate four values of the degenerate up and down quark masses $m_l$.
The bare masses are 0.015, 0.025, 0.035 and 0.050 in lattice units,
and cover a range of $M_\pi\!\sim\!290$\,--\,540~MeV.
The gauge ensembles are generated at a strange quark mass $m_s\!=\!0.080$,
close to its physical value $m_{s,\rm phys}\!=\!0.081$.
The $m_s$ dependence of the form factors is studied 
by calculating them at a different $m_s(=\!0.060)$ 
using the reweighting technique~\cite{reweight:1,reweight:2}.

The spatial lattice size is set to $N_s\!=\!L/a\!=\!24$ or 16 
depending on $m_l$,
so that a condition $M_\pi L \gtrsim 4$ is satisfied 
to control finite volume effects.
The temporal lattice size is fixed to $N_t \!=\! T/a \!=\! 48$. 
The statistics are 2,500 HMC trajectories at each simulation point $(m_l,m_s)$.
We estimate the statistical error by the jackknife method 
with a bin size of 50 trajectories.
Our simulation parameters are summarized in Table~\ref{tbl:method:param}.

\begin{ruledtabular}
\begin{table}[t]
\begin{center}
\caption{
   Simulation parameters. 
   Meson masses, $M_\pi$ and $M_K$, are in units of MeV.
}
\label{tbl:method:param}
\begin{tabular}{l|llll|llll}
   lattice               & $m_l$   & $m_s$   & $M_\pi$  & $M_K$  
                         & $\theta$ 
   \\ \hline
   $16^3 \!\times\! 48$  & 0.050   & 0.080   & 540(3)  & 617(4)
                         & 0.00, 0.40, 0.96, 1.60
   \\
   $16^3 \!\times\! 48$  & 0.035   & 0.080   & 453(4)  & 578(4)
                         & 0.00, 0.60, 1.28, 1.76
   \\
   $24^3 \!\times\! 48$  & 0.025   & 0.080   & 379(2)  & 548(3)
                         & 0.00, 1.68, 2.64
   \\
   $24^3 \!\times\! 48$  & 0.015   & 0.080   & 293(2)  & 518(1)
                         & 0.00, 1.68, 2.64
   \\ \hline
   $16^3 \!\times\! 48$  & 0.050   & 0.060   & 540(4)  & 567(4)
                         & 0.00, 0.40, 0.96, 1.60
   \\
   $16^3 \!\times\! 48$  & 0.035   & 0.060   & 451(6)  & 524(5)
                         & 0.00, 0.60, 1.28, 1.76
   \\
   $24^3 \!\times\! 48$  & 0.025   & 0.060   & 378(7)  & 492(7)
                         & 0.00, 1.68, 2.64
   \\
   $24^3 \!\times\! 48$  & 0.015  & 0.060    & 292(3)  & 459(4)
                         & 0.00, 1.68, 2.64
\end{tabular}
\end{center}
\vspace{0mm}
\end{table}
\end{ruledtabular}

\subsection{Calculation of meson correlators}


We calculate the three-point functions of the kaon and pion  
\bea
    C^{PQ}_{\mu,\phi\phi^\prime}
    (\dt, \dtp;\bfp,\bfp^\prime)
    & = &
    \frac{1}{N_t}
    \sum_{x_4=1}^{N_t}
    \sum_{\bfx,\bfxp,\bfxpp} 
    \langle 
       {\calO}_{Q,\phi^\prime}(\bfxpp,x_4+\dt+\dtp;\bfpp) 
    \nn \\
    &&
    \hspace{20mm}
       \times
       V_\mu(\bfxp,x_4+\dt)
       {\calO}_{P,\phi}(\bfx,x_4;\bfp)^{\dagger} 
    \rangle,
    \label{eqn:sim:msn_corr:msn_corr_3pt}  
\eea
where $P$ ($Q$) specifies the initial (final) meson,
and is ``$K$'' or ``$\pi$''.
The vector current $V_\mu$ is the weak current for $P\!\ne\!Q$, 
and light ($\bar{l}\gamma_\mu l$) or strange current ($\bar{s}\gamma_\mu s$)
for $P\!=\!Q$.
The initial (final) meson momentum is denoted by $\bfp^{(\prime)}$,
whereas 
$\dt^{(\prime)}$ is the temporal separation between $V_\mu$ 
and the meson source (sink) operator 
${\cal O}_{P,\phi}^\dagger$ (${\cal O}_{Q,\phi^{\prime}}$).
We also calculate the two-point function
\bea
    C^{P}_{\phi \phi^\prime}(\dt; \bfp)
    & =  &
    \frac{1}{N_t}
    \sum_{x_4=1}^{N_t}
    \sum_{\bfxp,\bfx} 
    \langle 
       {\calO}_{P,\phi^\prime}(\bfxp,x_4+\dt;\bfp) 
       {\calO}_{P,\phi}(\bfx,x_4;\bfp)^{\dagger} 
    \rangle,
    \label{eqn:sim:msn_corr:msn_corr_2pt} 
\eea
to extract the form factors 
below the maximum value of the momentum transfer
$\qsqmax\!=\!(M_K-M_\pi)^2$ (see Sec.~\ref{sec:ff} for details).


The meson interpolating field is given as 
\bea
   {\calO}_{P,\phi}(\bfx,x_4,\bfp)
   & = & 
   \sum_{\bfr}
   \phi(\bfr)\, 
   \bar{q}^\prime(\bfx+\bfr,x_4) \, 
   \gamma_5 \,
   q(\bfx,x_4).
   \label{eqn:sim:corr:msn_op} 
\eea
In addition to the simple local operator 
with $\phi_{l}(\bfr) \!=\! \delta_{\bfr,{\bf 0}}$, 
we also use an exponentially smeared operator 
with $\phi_{s}(\bfr) \!=\! \exp[-0.4|\bfr|]$
to reduce the excited state contamination 
in the meson correlation functions.


There is no explicit Fourier factors in the above expressions
(\ref{eqn:sim:msn_corr:msn_corr_3pt})
and (\ref{eqn:sim:msn_corr:msn_corr_2pt}).
The meson momentum $\bfp^{(\prime)}$ is induced 
through the twisted boundary condition for the valence quark fields~\cite{TBC} 
\bea
   q(\bfx+L \hat{k},x_4) 
   = 
   e^{i\theta} q(\bfx,x_4),
   \hspace{3mm}
   \bar{q}(\bfx+L \hat{k},x_4) 
   = 
   e^{-i\theta} \bar{q}(\bfx,x_4)
   \hspace{5mm}
   (k=1,2,3).
   \label{eqn:sim:msn_corr:tbc}
\eea
Here $\hat{k}$ is a unit vector in the $k$-th direction,
and we take a common twist angle $\theta$ in all three spatial directions
for simplicity.
This condition induces a quark momentum
of $p_k = \theta/L \leq 2\pi/L$,
and hence
a meson momentum 
$p_k\!=\!(\theta\!-\!\theta^\prime)/L$
by using different twist angles, $\theta$ and $\theta^{\prime}$, 
for the quark and anti-quark components.
With our choices of the twist angle listed in Table~\ref{tbl:method:param},
we simulate a region of the momentum transfer 
$-(300~\mbox{MeV})^2\!\lesssim\!t\!\leq\!\qsqmax$.
The important reference point $t\!=\!0$ is located inside this region.
Our studies of the EM form factors
of charged pion and kaon~\cite{EMFF:JLQCD:Nf3:RG+Ovr,PFF:JLQCD:Nf2:RG+Ovr} 
suggest that
the next-to-next-to-next-to leading order (N$^3$LO) chiral correction,
which is not known in ChPT, is small in this region of $t$.


In Eqs.~(\ref{eqn:sim:msn_corr:msn_corr_3pt})
and (\ref{eqn:sim:msn_corr:msn_corr_2pt}),
the summation over the source location $(\bfx,x_4)$ is not mandatory,
but is helpful to remarkably improve the  statistical accuracy.
To this end, we need the so-called all-to-all quark propagator,
which flows from any lattice site to any site.
Since a naive calculation is prohibitively time consuming,
we construct the all-to-all propagator by using 
low-lying modes of the overlap-Dirac operator~\cite{A2A:SESAM,A2A:TrinLat}
and the stochastic noise method~\cite{noise}.
Namely, 
the low-mode contribution is calculated as 
\bea
   \left\{D(m_q)^{-1}\right\}_{\rm low}(x,y)
   & = &
   \sum_{k=1}^{N_e} \frac{1}{\lambda_k^{(q)}} u_k(x) u_k^{\dagger}(y),
   \label{eqn:sim:a2a_prop:low}
\eea
where 
$\lambda_k^{(q)}$ and $u_k$ represent
the $k$-th lowest eigenvalue of $D(m_q)$ and its associated eigenvector.
The number of the low-modes is $N_e\!=\!240$ and 160
on the $24^3 \! \times \! 48$ and $16^3 \!\times\! 48$ lattices,
respectively.

\begin{figure}[t]
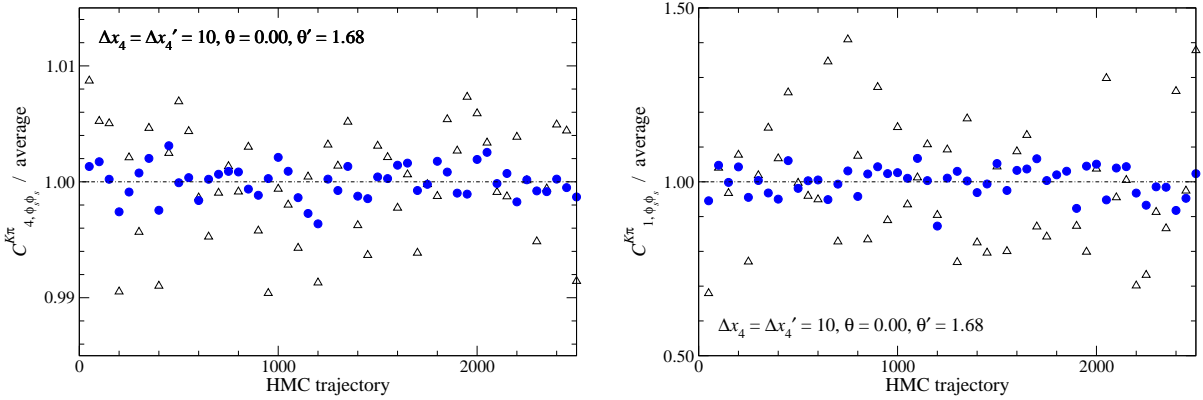

\begin{center}
   \includegraphics[angle=0,width=0.48\linewidth,clip]{jkd.p-p-v4_mud0_ms6_mval006_tbc010.dt10dtp10.eps}
   \hspace{3mm}
   \includegraphics[angle=0,width=0.48\linewidth,clip]{jkd.p-p-v1_mud0_ms6_mval006_tbc010.dt10dtp10.eps}

   \vspace{-3mm}
   \caption{
      Statistical fluctuation of three-point function 
      $C^{K\pi}_{\mu,\phi_s \phi_s}(\dt,\dtp;\bfp,\bfpp)$
      for $\mu\!=\!4$ (left panel) and 1 (right panel).
      We plot the value for each jackknife sample 
      normalized by the statistical average.
      The horizontal axis represents the HMC trajectory count 
      of the excluded configuration in the jackknife analysis.
      The data are obtained 
      at $(m_l,m_s)\!=\!(0.015,0.080)$
      with $\dt\!=\!\dtp\!=\!10$, $\theta\!=\!0.00$, $\theta^\prime\!=\!1.68$. 
      Triangles and circles are data before and after averaging over 
      the temporal location of the source operator $x_4$, respectively.
   }
   \label{fig:sim:msn_corr}
\end{center}
\end{figure}

We estimate the remaining contribution from higher modes
stochastically by using the noise method 
together with the dilution technique \cite{A2A:TrinLat}.
A complex $Z_2$ noise vector is prepared for each configuration,
and is split into $N_d = 3 \times 4 \times N_t/2$ vectors
$\eta_d(x) (d\!=\!1,\ldots,N_d)$. 
These diluted noise vectors have non-zero elements 
only for a single combination of 
color and spinor indices and at two consecutive time-slices.
We solve a linear equation for each diluted noise vector
\bea
   D(m_q)\,x_q
   = 
   P_{\rm high} \, \eta_d
   \hspace{5mm}           
   (d=1,\ldots,N_d),
   \hspace{2mm}           
   \label{eqn:sim:a2a_prop:high:leq}
\eea
where 
$P_{\rm high}\!=\!1-P_{\rm low}$, and 
$P_{\rm low} \!=\! \sum_{k=1}^{N_e} u_k\, u_k^{\dagger}$ 
is the projector to the eigenspace spanned by the low-modes. 
The high-mode contribution is then estimated as 
\bea
   \left\{ D(m_q)^{-1}\right\}_{\rm high}(x,y) 
   & = & 
   \sum_{d=1}^{N_d} x_d^{(q)}(x)\,\eta_d^{\dagger}(y).
   \label{eqn:sim:a2a_prop:high}
\eea
We refer the readers 
to Refs.~\cite{PFF:JLQCD:Nf2:RG+Ovr,EMFF:JLQCD:Nf3:RG+Ovr}
for more details on our implementation.


Figure~\ref{fig:sim:msn_corr} shows 
the statistical fluctuation of the three-point function
$C^{K\pi}_{\mu,\phi_s \phi_s}(\dt,\dtp;\bfp,\bfpp)$
with a certain choice of $\dt^{(\prime)}$ and $\bfp^{(\prime)}$.
Averaging over the temporal coordinate $x_4$ reduces the statistical error
by about a factor of 3 for the temporal component $\mu\!=\!4$ 
and a factor of 5 for spatial $\mu\!=\!1$.


The $m_s$ dependence of the form factors is studied 
by repeating our calculation at a different value of
the strange quark mass $m_s^\prime\!=\!0.060$
using the gauge ensemble at $m_s\!=\!0.040$
and the reweighing technique~\cite{reweight:1,reweight:2}.
For instance, the three-point function can be calculated as 
\bea
  \langle C^{PQ}_{\mu,\, \phi\phi^\prime} \rangle_{m_s^\prime}
  & = & 
  \langle C^{PQ}_{\mu,\, \phi\phi^\prime} \, \tilde{w}(m_s^\prime,m_s) \rangle_{m_s},
   \label{eq:rew}
\eea
where $\langle \cdots \rangle_{m_s^{(\prime)}}$ represents the Monte Carlo average 
at $m_s^{(\prime)}$. 
The reweighting factor $\tilde{w}$ from $m_s$ to $m_s^\prime$
is defined as 
\bea
   \tilde{w}(m_s^\prime,m_s)
   & = & 
   \frac{w(m_s^\prime,m_s)}{\langle w(m_s^\prime,m_s) \rangle_{m_s}},
   \hspace{5mm}
   w(m_s^\prime,m_s)
   = 
   \det\left[ \frac{D(m_s^\prime)}{D(m_s)} \right]
   \label{eqn:rw_fctr}
\eea
for each gauge configuration.
In order to remarkably reduce the computational cost, 
$w$ is decomposed into contributions from low- and high-modes
\bea
   w(m_s^\prime,m_s)
   & = & 
   w_{\rm low}(m_s^\prime,m_s) \,w_{\rm high}(m_s^\prime,m_s),
   \\
   w_{\rm low (high)}(m_s^\prime,m_s) 
   & = & 
   \det\left[ 
      P_{\rm low (high)} 
      \frac{D(m_s^\prime)}{D(m_s)} 
      P_{\rm low (high)}
   \right].
\eea
We exactly calculate the low mode contribution $w_{\rm low}$ 
by using the low-lying eigenvalues, 
whereas the high-mode contribution $w_{\rm high}$ 
is estimated through a stochastic estimator for its square 
\begin{eqnarray}
 w_{\rm{high}}^2(m_s^\prime,m_s)
  =
  \frac{1}{N_r}
  \sum_{r=1}^{N_r} e^{-\frac{1}{2}(P_{\rm high}\xi_r)^\dagger 
                (\Omega-1) P_{\rm high} \xi_r},
  \label{eq:high}
\end{eqnarray}
where
$\Omega 
 \equiv D(m_s)^\dagger \{D(m_s^\prime)^{-1}\}^\dagger D(m_s^\prime)^{-1}D(m_s)$.
In our study of the EM form factors~\cite{EMFF:JLQCD:Nf3:RG+Ovr},
the full reweighting factor $\omega$ turned out to be largely dominated 
by the low-mode contribution $w_{\rm low}$ with our simulation set-up.
For each configuration, therefore, 
we use only ten Gaussian random vectors
$\{\xi_1,...,\xi_{10}\}$ ($N_r\!=\!10$) 
with which the uncertainty of $\tilde{w}$ due to the stochastic estimator 
is negligibly small compared to its statistical fluctuation.

\section{Form factors at simulation points}
\label{sec:ff}


\subsection{form factors}


\begin{figure}[t]
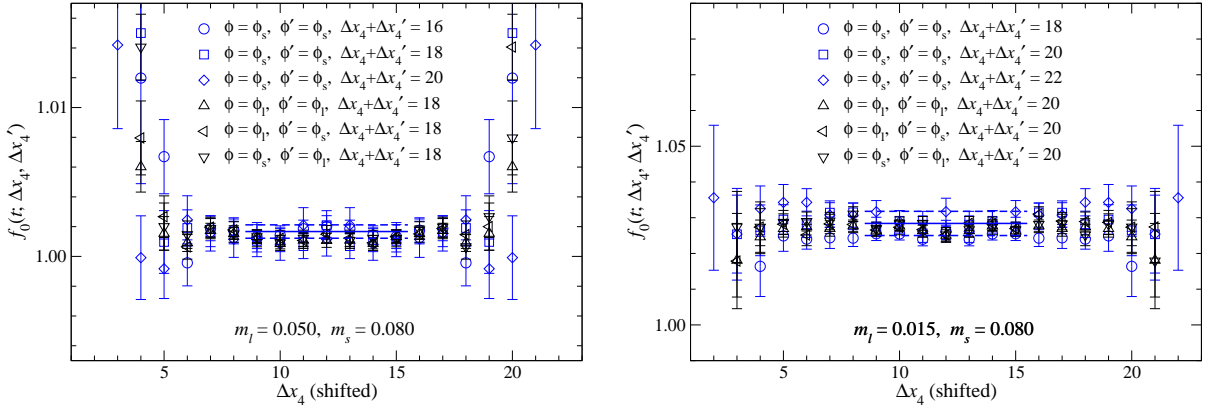

\begin{center}
   \includegraphics[angle=0,width=0.48\linewidth,clip]%
                   {kff_drat14_vs_dtsrc.conv.mud3_ms6.eps}
   \hspace{3mm}
   \includegraphics[angle=0,width=0.48\linewidth,clip]%
                   {kff_drat14_vs_dtsrc.conv.mud0_ms6.eps}

   \vspace{0mm}
   \caption{
      Effective value of $f_0(t_{\rm max})$ estimated from double ratio
      $R_{\phi\phi^\prime}(\dt,\dtp)$ 
      at $(m_l,m_s)\!=\!(0.050,0.080)$ (left panel) 
      and (0.015,0.080) (right panel).
      Blue circles, squares and diamond show
      data with the smeared source and sink
      for different values of $\dt+\dtp$.
      On the other hand,       
      black triangles show data with local source and/or sink
      with $\dt+\dtp$ kept fixed.
      All data are shifted along the horizontal axis 
      so that the meson source and sink operators are located 
      at $T/4-(\dt+\dtp)/2$ and $T/4+(\dt+\dtp)/2$,
      respectively.
      Solid and dashed lines show a constant fit to data 
      with different values of $\dt$ and $\dt+\dtp$.
   }
   \label{fig:ff:f0tmax_vs_dtsrc}
\end{center}
\vspace{0mm}
\end{figure}

In the limit of large temporal separations $\dt,\dtp\!\to\!\infty$,
the light meson three-point function ($P,Q\!=\!\pi$ or $K$)
is dominated by the ground state contribution as 
\bea
   C^{PQ}_{\mu, \phi\phi^\prime}(\dt, \dtp;\bfp,\bfpp)
   & \xrightarrow[\dt,\dtp \to \infty]{} &
   \frac{Z_{Q,\phi^\prime}(\bfpp)^*\, Z_{P, \phi}(\bfp)} 
   {4\, E_Q(\bfpp) E_P(\bfp)}
   \frac{1}{Z_V}
   \langle Q(p^\prime) | V_\mu | P(p) \rangle\,
   \nn \\
   &   &
   \hspace{25mm} 
   \times 
   e^{-E_Q(\bfpp)\, \dtp} e^{-E_P(\bfp)\, \dt},
   \label{eqn:ff:ff:msn_corr_3pt}
\eea
where 
$Z_{P,\phi}(\bfp)\!=\!\langle P(p) | {\calO}_{P,\phi} \rangle$ 
is the overlap of the meson interpolating field to the physical state,
and $Z_V$ is the renormalization factor for the vector current.
These factors and the exponential damping factors 
$e^{-E_{P(Q)}(\bfp^{(\prime)})\dt^{(\prime)}}$
cancel in the following double ratio~\cite{dble_ratio,Kl3:Rome:Nf0:Plq+Clv}
\bea
   R_{\phi\phi^\prime}(\dt, \dt^\prime) 
   & = & 
   \frac{C^{K \pi}_{4,\phi\phi^\prime}(\dt,\dtp; \bfz, \bfz)
         C^{\pi K}_{4,\phi\phi^\prime}(\dt,\dtp; \bfz, \bfz)}
        {C^{K K}_{4,\phi\phi^\prime}(\dt,\dtp; \bfz, \bfz)
         C^{\pi \pi}_{4,\phi\phi^\prime}(\dt,\dtp; \bfz, \bfz)}
   \nn \\
   & & 
   \xrightarrow[\dt, \dtp \to \infty]{} 
   \frac{(M_K+M_\pi)^2}{4 M_K M_\pi} f_0(t_{\rm max})^2,
   \label{eqn:ff:drat1}
\eea
from which we calculate the scalar form factor $f_0(t_{\rm max})$ 
at the largest momentum transfer $t_{\rm max}$.
Figure~\ref{fig:ff:f0tmax_vs_dtsrc} shows 
the effective value of $f_0(t_{\rm max})$ as a function of $\dt$.
The accuracy of $f_0(t_{\rm max})$ is typically $\lesssim\!1$\,\%
with our simulation set-up.
The figure also demonstrates that
the all-to-all quark propagator greatly helps us 
increase the reliability of the precision calculation of $f_0(t_{\rm max})$:
it enables us to confirm the consistency in $f_0(\qsqmax)$
among different values of $\dt+\dtp$
and different smearing functions for the meson interpolating fields.


At smaller momentum transfer $t < t_{\rm max}$,
the vector form factor $f_+(t)$ and the ratio $\xi(t)=f_-(t)/f_+(t)$
are calculated from the following ratios~\cite{Kl3:Rome:Nf0:Plq+Clv,Kl3:JLQCD:Nf2:Plq+Clv,Kl3:RBC:Nf2:DBW2+DWF}
\bea
   \tilde{R}_{\phi\phi^\prime}(\bfp,\bfpp; \dt, \dtp) 
   = 
   \frac{C^{K \pi}_{4,\phi\phi^\prime}(\dt,\dtp; \bfp, \bfpp)
         C^{K}_{\phi\phi_l}(\dt, \bfz)\, C^{\pi}_{\phi_l \phi^\prime}(\dtp, \bfz)}
        {C^{K \pi}_{4,\phi\phi^\prime}(\dt, \dtp; \bfz, \bfz)
         C^{K}_{\phi\phi_l}(\dt,\bfp)\, C^{\pi}_{\phi_l \phi^\prime}(\dtp, \bfpp)}
   \nn \\
   \xrightarrow[\dt, \dtp \to \infty]{} 
   \left\{
      \frac{E_K(\bfp)+E_\pi(\bfpp)}{M_K+M_\pi} 
    + \frac{E_K(\bfp)-E_\pi(\bfpp)}{M_K+M_\pi} \xi(t)
   \right\}
   \frac{f_+(t)}{f_0(t_{\rm max})},
   \hspace{9mm}
   \label{eqn:ff:drat2}
   \\[1mm]
   R_{k,\phi\phi^\prime}(\bfp,\bfpp; \dt, \dtp)
   = 
   \frac{C^{K \pi}_{k,\phi\phi^\prime}(\dt,\dtp; \bfp, \bfpp)
         C^{KK}_{4,\phi\phi^\prime}(\dt, \dtp; \bfp, \bfpp)}
        {C^{K \pi}_{4,\phi\phi^\prime}(\dt, \dtp; \bfp, \bfpp)
         C^{KK}_{k,\phi\phi^\prime}(\dt, \dtp; \bfp, \bfpp)}
   \nn \\
   \hspace{5mm} \xrightarrow[\dt, \dtp \to \infty]{} 
   \frac{E_K(\bfp)+E_K(\bfpp)}{(p+\pp)_k}
   \frac{(p+\pp)_k + (p-\pp)_k \xi(t)}
        {E_K(\bfp)+E_\pi(\bfpp) + \left\{E_K(\bfp)-E_\pi(\bfpp)\right\}\xi(t)}.
   \label{eqn:ff:drat3}
\eea
The last line of Eq.~(\ref{eqn:ff:drat2})
assumes the asymptotic form of the two-point function
\bea
   C^{P}_{\phi\phi^\prime}(\dt;\bfp)
   & \xrightarrow[\dt \to \infty]{} &
   \frac{Z_{P,\phi^\prime}(\bfp)^*\, Z_{P, \phi}(\bfp)} 
        {2\, E_P(\bfp)}\,
   e^{-E_P(\bfp)\,\dt}.
   \label{eqn:ff:ff:msn_corr_2pt}
\eea
We evaluate $f_0(t)$ from $f_+(t)$ and $\xi(t)$ at $t < t_{\rm max}$. 
Note that, at $t_{\rm max}$, we only have results for $f_0(t_{\rm max})$ 
from $R_{\phi\phi^\prime}$,
since $\tilde{R}_{\phi\phi^\prime}$ and $R_{k,\phi\phi^\prime}$ 
have no sensitivity to $f_+$ and $\xi$.


\begin{figure}[t]
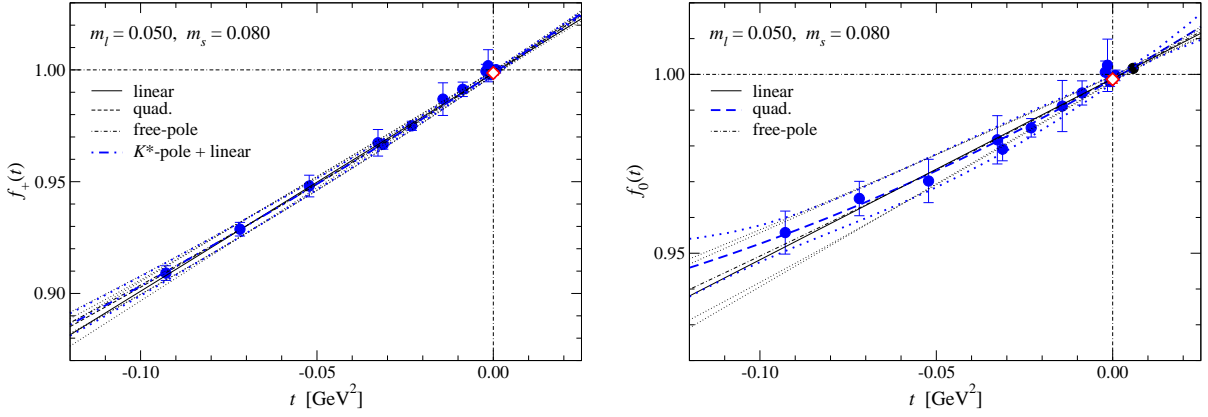

\begin{center}
\includegraphics[angle=0,width=0.48\linewidth,clip]%
                {./f+_vs_q2_mud0050_ms0080.phys.eps}
\hspace{3mm}
\includegraphics[angle=0,width=0.48\linewidth,clip]%
                {./f0_vs_q2_mud0050_ms0080.phys.eps}

\vspace{-3mm}
\caption{
   Vector (left panel) and scalar form factors (right panel)
   as a function of $t$ at $(m_l,m_s)\!=\!(0.050,0.080)$.
   Solid circles show the results at simulated $t$'s. 
   Solid, dashed and dot-dashed lines are linear, quadratic
   (\ref{eqn:ff:t-dep:poly}) 
   and free-pole fits (\ref{eqn:ff:t-dep:f-pole}), respectively. 
   We also plot the fit with the $K^*$ pole 
   (\ref{eqn:ff:t-dep:vmd+poly}) for $f_+$. 
   The dotted lines show their statistical error.
   The value interpolated to $t\!=\!0$ is shown by the open diamond.
   The blue thicker lines show the fit to estimate the central value 
   and statistical error of the interpolated value.
}
\label{fig:ff:kff_vs_q2:mud0050_ms0080}
\end{center}
\end{figure}

\begin{figure}[t]
\begin{center}
\includegraphics[angle=0,width=0.48\linewidth,clip]%
                {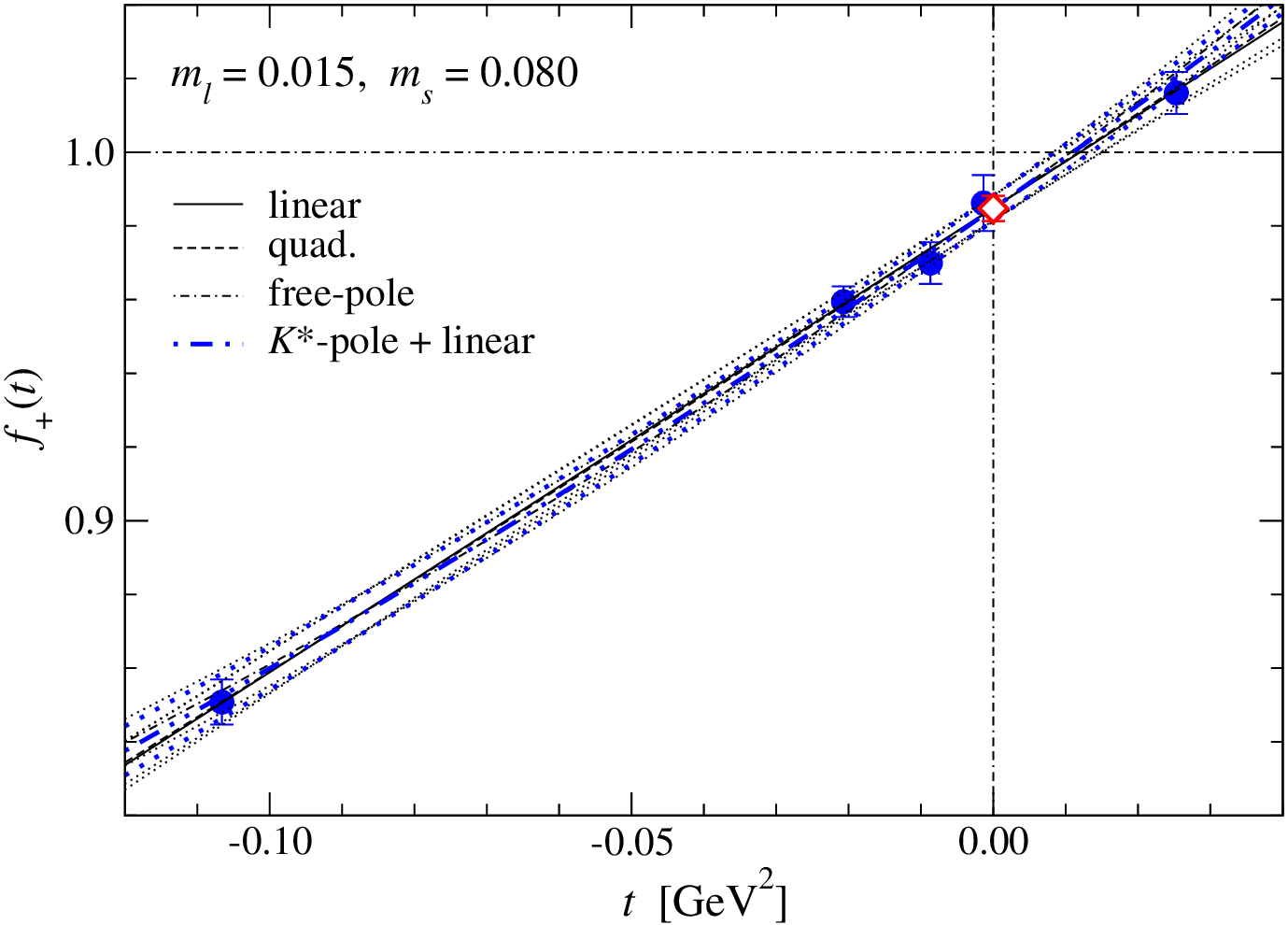}
\hspace{3mm}
\includegraphics[angle=0,width=0.48\linewidth,clip]%
                {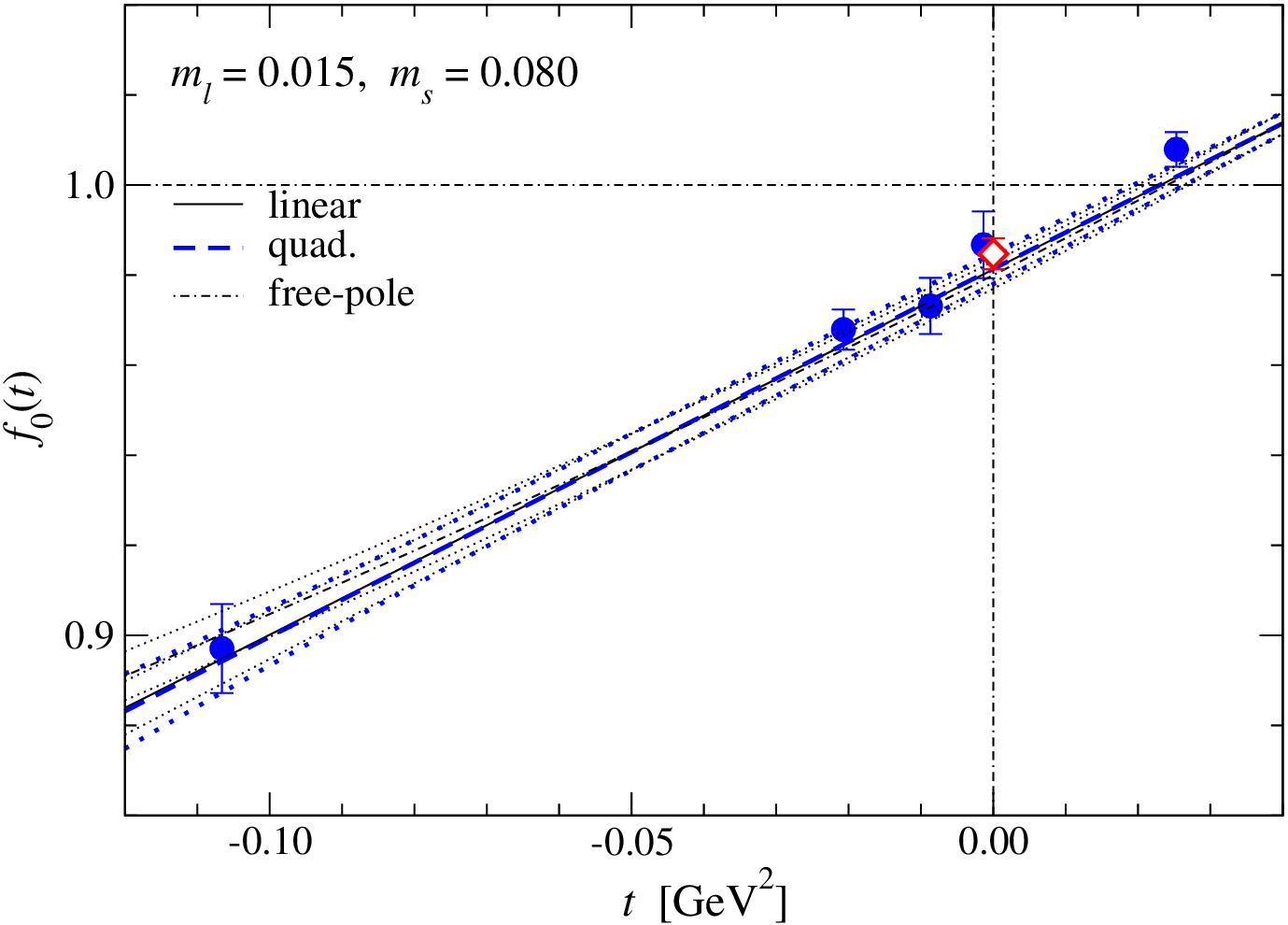}

\vspace{-3mm}
\caption{
   Same as Fig.~\protect\ref{fig:ff:kff_vs_q2:mud0050_ms0080},
   but for $(m_l,m_s)\!=\!(0.015,0.080)$.
}
\label{fig:ff:kff_vs_q2:mud0015_ms0080}
\end{center}
\end{figure}

\begin{figure}[t]
\begin{center}
\includegraphics[angle=0,width=0.48\linewidth,clip]%
                {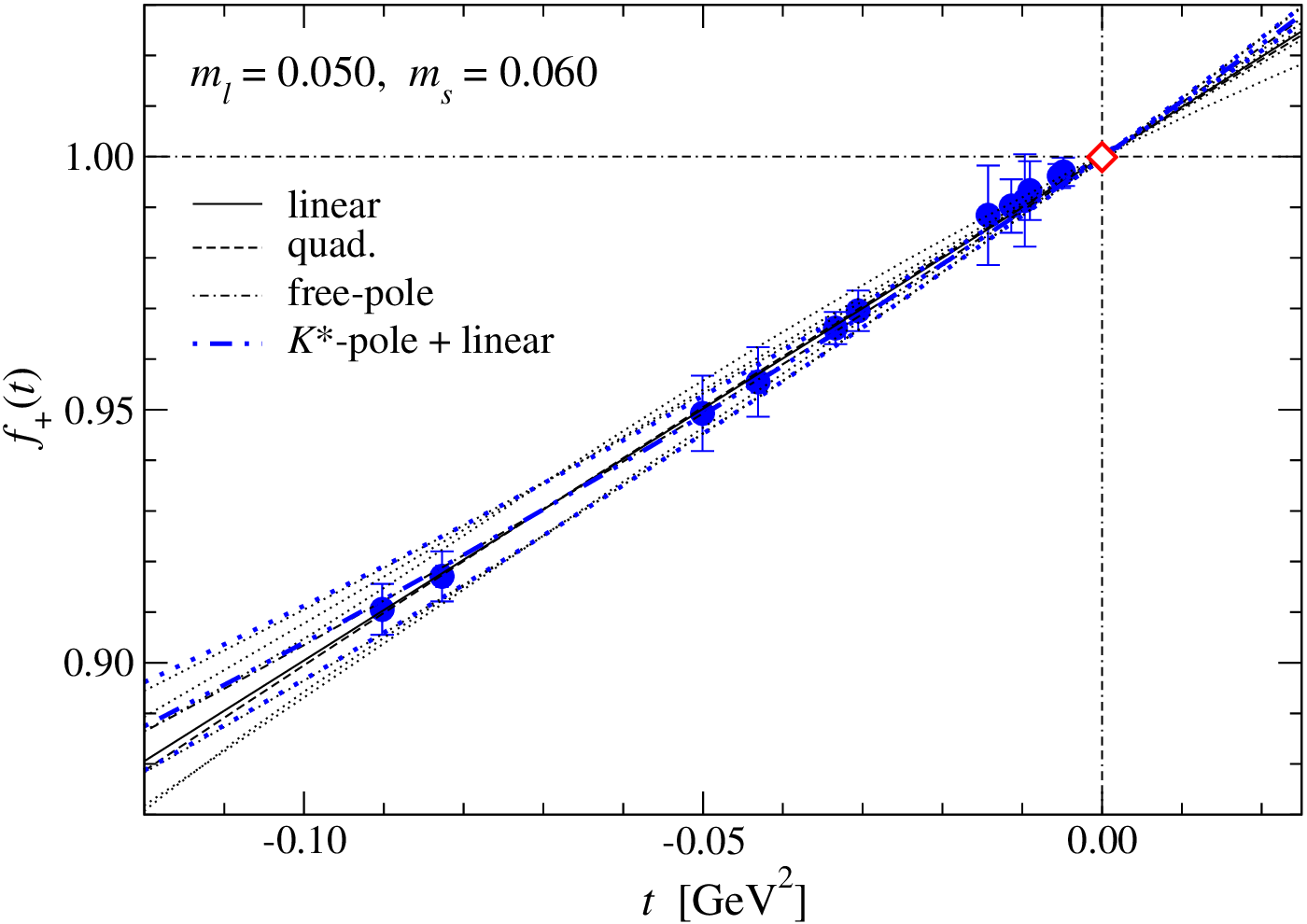}
\hspace{3mm}
\includegraphics[angle=0,width=0.48\linewidth,clip]%
                {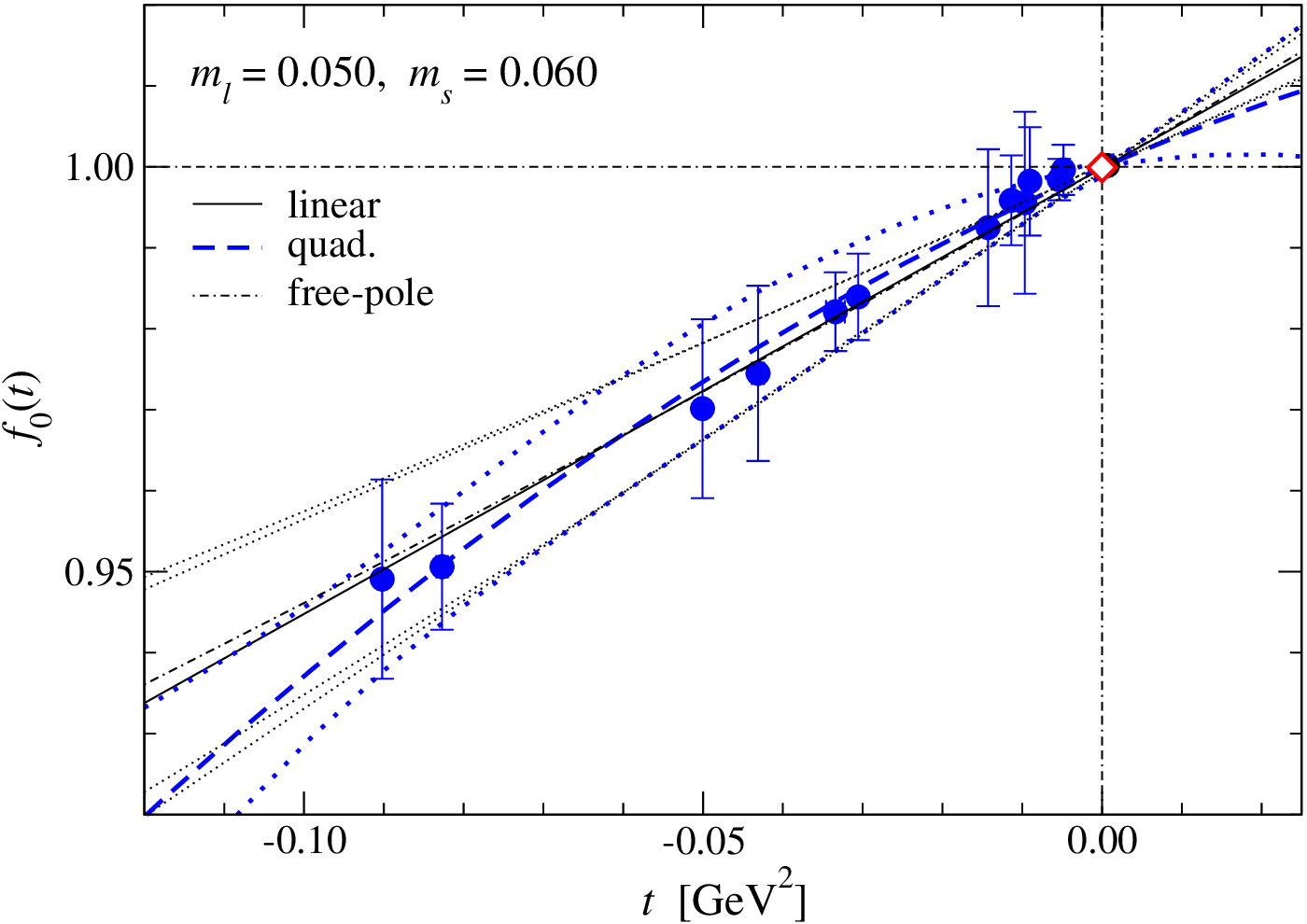}

\vspace{-3mm}
\caption{
   Same as Fig.~\protect\ref{fig:ff:kff_vs_q2:mud0050_ms0080},
   but for $(m_l,m_s)\!=\!(0.050,0.060)$.
}
\label{fig:ff:kff_vs_q2:mud0050_ms0060}
\end{center}
\end{figure}

In Figs.~\ref{fig:ff:kff_vs_q2:mud0050_ms0080} 
and \ref{fig:ff:kff_vs_q2:mud0015_ms0080},
we plot the vector and scalar form factors as a function of $t$.
At $m_s\!=\!0.080$,
the statistical accuracy of the non-trivial chiral correction $f_{\{+,0\}}(t)-1$ 
due to $m_s\!\ne\!m_l$ and $t\!\ne\!0$ is typically 
10\,--\,20\,\%.
Analyses based on ChPT suggest that 
finite volume effects including 
those due to the twisted boundary condition~\cite{FVE:ChPT:TBC:BR}
are below this accuracy: 
$O(\exp[ -M_\pi L ])\!\lesssim\!O(2\,\%)$ or less~\cite{FVE:ChPT:TBC:GS,FVE:ChPT:TBC:BR,FVE:ChPT:TBC:BBGR}.

In Fig.~\ref{fig:ff:kff_vs_q2:mud0050_ms0060},
we observe that
the statistical error is about a factor of two larger 
at $m_s\!=\!0.060$ due to reweighting.


We parametrize the $t$ dependence of the form factors $f_{\{+,0\}}$
to estimate the normalization $f_+(0)\!=\!f_0(0)$ and 
the slopes $d f_{\{+,0\}}/dt|_{t=0}$ at simulated quark masses.
For the vector form factor $f_+$, 
we use the following parametrization 
based on the vector meson dominance (VMD) hypothesis
\bea
   f_+(t)
   & = &
   f_+(0) 
   \left\{
      \frac{1}{1-t/M_{K^*}^2} + a t
   \right\},
   \label{eqn:ff:t-dep:vmd+poly}
\eea
where $M_{K^*}$ represents the strange-light vector meson mass 
calculated at simulation points.
There could be contribution of the higher poles and cuts,
which is well approximated by the additional linear term 
at our small values of $|t|$.
We note that 
our data of the pion and kaon EM form factors
in a similar region of $t$
are well described by this type of the parametrization~\cite{EMFF:JLQCD:Nf3:RG+Ovr}. 

Since we simulate small values of $|t|$, 
our data of $f_{\{+,0\}}$ does not show any strong curvature 
in Figs.~\ref{fig:ff:kff_vs_q2:mud0050_ms0080}\,--\,\ref{fig:ff:kff_vs_q2:mud0050_ms0060}, 
and simple polynomial parametrization also describes our data well
\bea
   f_{\{+,0\}}(t)
   & = &
   f_+(0)
   \left\{
      1 
    + \lambdapzp\,\frac{t}{M_{\pi^\pm,\rm phys}^2} 
    + \lambda_{\{+,0\}}^{\prime\prime} 
      \left( \frac{t}{M_{\pi^\pm,\rm phys}^2} \right)^2
   \right\}
   \label{eqn:ff:t-dep:poly}
\eea
even without the quadratic term.
The conventional free-pole form
\bea
   f_{\{+,0\}}(t)   
   & = &
   \frac{f_{\{+,0\}}(0)}{1-t/M_{{\rm pole},\{+,0\}}^2}
   \label{eqn:ff:t-dep:f-pole}
\eea
also works well.

\begin{ruledtabular}
\begin{table}[t]
\begin{center}
\caption{
   Fit results for normalization $f_+(0)$ 
   and slope parameters $\lambda_{\{+,0\}}^\prime$.
}
\label{tbl:ff:q2_interp}
\begin{tabular}{ll|llllll}
   $m_l$  & $m_s$  & $\fp(0)$ & $\lambda_+^\prime \!\times\! 10^2$ 
                              & $\lambda_0^\prime \!\times\! 10^2$ 
   \\ \hline
   0.050  & 0.080  &  $0.9986(11)\left(^{+2}_{-1}\right)$
                   &  $2.02(7)(12)$
                   &  $1.10(28)(11)$
   \\
   0.050  & 0.060  &  $0.99991(25)(-8)$
                   &  $2.12(14)(22)$
                   &  $0.84(51)(27)$
   \\ \hline
   0.035  & 0.080  &  $0.9937(31)(+8)$
                   &  $2.37(11)(18)$
                   &  $1.42(29)(11)$
   \\
   0.035  & 0.060  &  $0.9977(20)\left(^{+4}_{-2}\right)$
                   &  $2.52(19)(19)$
                   &  $1.72(42)(26)$
   \\ \hline
   0.025  & 0.080  &  $0.9919(18)\left(^{+7}_{-5}\right)$
                   &  $2.51(5)(21)$
                   &  $1.46(17)(7)$
   \\
   0.025  & 0.060  &  $0.9969(46)\left(^{+1}_{-11}\right)$
                   &  $2.74(14)(26)$
                   &  $1.59(52)(18)$
   \\ \hline
   0.015  & 0.080  &  $0.9847(34)\left(^{+4}_{-5}\right)$
                   &  $2.72(10)(22)$
                   &  $1.70(15)(1)$
   \\
   0.015  & 0.060  &  $0.9922(74)\left(^{+18}_{-7}\right)$
                   &  $2.75(13)(32)$
                   &  $1.92(32)(14)$
\end{tabular}
\end{center}
\vspace{5mm}
\end{table}
\end{ruledtabular}

In this study,
we estimate $f_+(0)$ and $df_{\{+,0\}}/dt|_{t=0}$ by a simultaneous fit 
using the VMD-based form~(\ref{eqn:ff:t-dep:vmd+poly}) for $f_+$
and the quadratic form~(\ref{eqn:ff:t-dep:poly}) for $f_0$.
The uncertainty due to the choice of the fitting form is estimated by testing 
the polynomial and free-pole forms for $f_+$
and the linear and free-pole forms for $f_0$.
Fit results are summarized in Table~\ref{tbl:ff:q2_interp},
where we list the phenomenologically familiar slope parameter
$\lambdapzp$ 
in the quadratic parametrization (\ref{eqn:ff:t-dep:poly})
instead of 
$df_{\{+,0\}}/dt|_{t=0}\!=\!f_+(0) \lambdapzp / M_{\pi^\pm,\rm phys}^2$.

In the simulated region of $t$,
all the aforementioned parametrizations describe our data well 
with $\chi^2/{\rm d.o.f}\!\lesssim\!0.5$.
The choice of the parametrization leads to small uncertainty for $f_+(0)$
compared to the statistical accuracy. 
For the slope parameter $\lambda_{\{+,0\}}^\prime$, 
the systematic error is more important,
but not so large compared to the statistical one.

\section{Chiral extrapolation of form factors}
\label{sec:chiral_fit}


\subsection{ChPT formulae and LECs}


The momentum transfer and quark mass dependence of the form factors 
is known up to NNLO in SU(3) ChPT~\cite{KFF:weak:ChPT:SU3:NNLO:PS,KFF:weak:ChPT:SU3:NNLO:BT}.
Let us denote the chiral expansion as
\bea
   f_X(t)
   & = &
   f_{X,0} + f_{X,2}(t) + f_{X,4}(t) + f_{X,6}(t) 
   \hspace{3mm}
   (X=+,-,0),
   \label{eqn:chiral_fit:ff}
\eea
where $f_{X,0}$, $f_{X,2}$, and $f_{X,4}$, represent 
the LO, NLO, and NNLO contributions, respectively.
We add a possible higher order term $f_{X,6}$,
the functional form of which is not yet known.


The current conservation fixes the normalization of the vector form factor
in the chiral limit as $\fplo\!=\!1$.
The NLO contribution can be decomposed into two parts
\bea
   \fpnlo(t)
   & = &
   \fpnloL(t) + \fpnloB(t).
   \label{eqn:chiral_fit:fp:nlo}
\eea
The first part $\fpnloL$ represents 
the analytic term arising from the tree diagram
with a vertex arising from the $O(p^4)$ chiral Lagrangian ${\mathcal L}_4$,
\bea
   F_\pi^2\, \fpnloL(t)
   & = & 
   2 L_9^r t.
   \label{eqn:chiral_fit:fp:nlo_l}
\eea
Note that $p$ symbolically represents the Nambu-Goldstone (NG) boson momentum,
and $L_9^r$ is a LEC in ${\mathcal L}_4$.
This contribution does not involve quark masses to be compatible
with the current conservation $\fpnloL(0)\!=\!0$.


The other part is the contribution of loop diagrams
\bea
   F_\pi^2\, \fpnloB(t)
   & = & 
   \frac{3}{8} 
   \left\{ 
      \bar{A}(M_\pi^2) + 2 \bar{A}(M_K^2) + \bar{A}(M_\eta^2) 
   \right\}
   \nn
   \\
   && 
   \hspace{10mm}
  -\frac{3}{2}
   \left\{
      \bar{B}_{22}(M_\pi^2,M_K^2,t) + \bar{B}_{22}(M_K^2,M_\eta^2,t) 
   \right\},
   \label{eqn:chiral_fit:fp:nlo_b}
\eea   
where $\bar{A}$ and $\bar{B}_{22}$ represent one-loop integral functions.
We refer the readers 
to Refs.~\cite{EMFF:JLQCD:Nf3:RG+Ovr,PFF+KFF:ChPT:NNLO:Nf3}
for their definition and expression.
Note that
the so-called $\xi$-expansion is employed in this study: 
the form factors are expanded 
in terms of $\xi_{\{\pi,K,\eta\}}\!=\!M_{\{\pi,K,\eta\}}^2/(4\pi F_\pi)^2$,
where $F_\pi$ represents the pion decay constant.
In Ref.~\cite{Spectrum:Nf2:RG+Ovr:JLQCD},
we demonstrated that 
the $\xi$-expansion of the meson masses and decay constants 
has a better convergence than 
the expansion in terms of $m_q/(4\pi F_0)^2$.
Here $F_0$ is the decay constant in the chiral limit. 
The $\xi$-expansion has another important advantage that 
ChPT formulae are free from the unknown LEC $F_0$.


This is also the case for the NNLO contribution
\bea
   \fpnnlo(t)
   & = &
   \fpnnloC(t) + \fpnnloL(t) + \fpnnloB(t).
   \label{eqn:chiral_fit:fp:nnlo}
\eea
Here $\fpnnloB$ represents the contribution of two-loop diagrams
without any vertices from ${\mathcal L}_4$
and $O(p^6)$ chiral Lagrangians ${\mathcal L}_6$~\cite{ChPT:SU3:NNLO}. 
While its expression is rather lengthy~\cite{KFF:weak:ChPT:SU3:NNLO:BT:loop},
it does not contain any LECs in the $\xi$\,-\,expansion,
and is not an obstacle to obtaining a stable chiral extrapolation.


The term $\fpnnloL$ mainly arises from the one-loop diagrams
with one vertex from $\mathcal{L}_4$,
and hence depends on the $O(p^4)$ couplings $L_i^r$.
At the level of NLO, 
only $L_9^r$ appears in $\fpnloL$,
and we fix it to an estimate obtained from our study of the EM form 
factors~\cite{EMFF:JLQCD:Nf3:RG+Ovr}.
Other $L_{\{1\,-\,8\}}$ appear only in small NNLO term $\fpnnloL$. 
We fix them to a recent phenomenological estimate 
in Ref.~\cite{ChPT:LECs:SU2+SU3}.
These input values are listed in Table~\ref{tbl:chiral_fit:input:Li}.

\begin{ruledtabular}
\begin{table}[t]
\begin{center}
\caption{
   Input values for $O(p^4)$ couplings $L_i^r$
   at renormalization scale $\mu\!=\!M_\rho$.
   We use $L_9^r$ from our study of the EM form factors~\cite{EMFF:JLQCD:Nf3:RG+Ovr},
   whereas $L_{\{1\,-\,8\}}$ are taken from a phenomenological study~\cite{ChPT:LECs:SU2+SU3}. 
   In that paper, authors presented two estimates obtained from 
   different ChPT fits of experimental data.
   The central value and the first statistical error of $L_{\{1\,-\,8\}}$ are
   from the authors' preferred fit, 
   whereas we assign the difference between the two estimates 
   as the second systematic error.
}
\label{tbl:chiral_fit:input:Li}
\begin{tabular}{lllll}
   $L_1^r\!\times\!10^3$   & $L_2^r\!\times\!10^3$   & 
   $L_3^r\!\times\!10^3$   & $L_4^r\!\times\!10^3$   & 
   $L_5^r\!\times\!10^3$   
   \\ \hline
   0.53(6)(+11)  &  0.81(4)(-22)  & -3.07(20)(+27)  & 0.3(0)(+0.46)  &  
   1.01(6)(-51)
   \\ \hline
   $L_6^r\!\times\!10^3$   & $L_7^r\!\times\!10^3$   & 
   $L_8^r\!\times\!10^3$   & $L_9^r\!\times\!10^3$   
   \\ \hline
   0.14(5)(+35)  & -0.34(9)(+15)  &  0.47(10)(-30)  & 
   $4.6(1.1)\left(^{+0.1}_{-0.5}\right)$
\end{tabular}
\end{center}
\vspace{0mm}
\end{table}
\end{ruledtabular}


The NNLO analytic term $\fpnnloC$ arises from tree-diagrams with 
one vertex from $\mathcal{L}_6$. 
A central issue in our analysis based on NNLO ChPT is 
how to deal with many $O(p^6)$ couplings $C_i^r$ appearing 
in this contribution
\bea
   F_\pi^4 \fpnnloC
   & = & 
  -8 \Cplspk\,(M_K^2-M_\pi^2)^2
  -4 \Cplspt\,M_\pi^2\,t
  -4 \Cplskt\,M_K^2\,t
  -4 \Ctt\, t^2,
   \label{eqn:chiral_fit:fp:nnlo_c}
\eea
where the coefficients $c^r_X$'s are
linear combinations of $O(p^6)$ couplings
\bea
   \Cplspk
   & = & 
   C_{12}^r + C_{34}^r,
   \label{eqn:chiral_fit:f+:Cplspk}
   \\
   c_{+,\pi t}^r
   & = & 
   2 C_{12}^r + 4 C_{13}^r + C_{64}^r + C_{65}^r + C_{90}^r,
   \\
   c_{+,K t}^r
   & = &
   2 C_{12}^r + 8 C_{13}^r + 2 C_{63}^r + 2C_{64}^r + C_{90}^r,
   \\
   \Ctt
   & = &
   C_{88}^r - C_{90}^r.
   \label{eqn:chiral_fit:f+:Ctt}
\eea
Similar to the case of $L_i^r$,
the chiral behavior of the EM form factors provides helpful
information about $C_i^r$.
The coefficient of the $O(t^2)$ term, namely $\Ctt$, is the same 
as the NNLO analytic term of the EM form factors
\bea
   F_\pi^4\,
   \pffnnloC(t)
   & = & 
   -4 \Cppt\, M_\pi^2\, t  - 8 \Cpkt\, M_K^2\, t  - 4\Ctt\, t^2,
   \label{eqn:chiral_fit:pff:nnlo_c}
   \\
   F_\pi^4\,
   \kpffnnloC(t)
   & = &
   -4 \Ckpt\, M_\pi^2\, t  - 4 \Ckkt\, M_K^2\, t  - 4\Ctt\, t^2,
   \label{eqn:chiral_fit:kpff:nnlo_c}
   \\
   F_\pi^4\,
   \knffnnloC(t)
   & = &
  -\frac{8}{3}\, \Ckn\, (M_\pi^2 - M_K^2 )\, t,
   \label{eqn:chiral_fit:knff:nnlo_c}
\eea 
where 
\bea
   \Cppt 
   & = & 
   4 C_{12}^r + 4 C_{13}^r + 2 C_{63}^r + C_{64}^r + C_{65}^r + 2 C_{90}^r, 
   \\
   \Cpkt
   & = & 
   4 C_{13}^r + C_{64}^r,
   \\
   \Ckpt 
   & = &
   4 C_{13}^r + \frac{2}{3} C_{63}^r + C_{64}^r - \frac{1}{3} C_{65}^r, 
   \\
   \Ckkt
   & = &     
   4 C_{12}^r + 8 C_{13}^r + \frac{4}{3} C_{63}^r + 2 C_{64}^r 
                 + \frac{4}{3} C_{65}^r + 2 C_{90}^r,
   \\
   \Ckn
   & = &
   2C_{63}^r - C_{65}^r.
\eea 
In addition, 
two coefficients for the $K\!\to\!\pi$ decays, $\Cplspt$ and $\Cplskt$, 
are written in terms of 
those for the EM form factors as 
\bea
  c_{+,\pi t}^r
  & = & 
  \frac{1}{2}\left( \Cppt + \Cpkt - \Ckn \right),
  \\
  c_{+,K t}^r
  & = &
  \frac{1}{2}\left( \Cppt + 3 \Cpkt + \Ckn \right).
\eea
Therefore, we have only single free parameter $\Cplspk$
in our chiral extrapolation of $f_+$ at the level of NNLO.
The term $-8\Cplspk(M_K^2-M_\pi^2)^2$ in $\fpnnloC$ describes 
the SU(3) breaking effects at $t\!=\!0$,
and hence is absent in the EM form factors. 
The coefficient $\Cplspk$ is therefore to be determined from the data of $f_+$.
For other coefficients $\Cplspt$, $\Cplspt$ and $\Ctt$, 
we use our estimate~\cite{EMFF:JLQCD:Nf3:RG+Ovr},
which is summarized in Table~\ref{tbl:chiral_fit:input:Ci}.
The uncertainty due to the choice of the input 
in Tables~\ref{tbl:chiral_fit:input:Li} and \ref{tbl:chiral_fit:input:Ci}
is estimated by repeating the following analysis 
with the input shifted by its uncertainty quoted in the tables.

\begin{ruledtabular}
\begin{table}[t]
\begin{center}
\caption{
   Input values for the linear combinations of $O(p^6)$ couplings
   obtained in our study of EM form factors~\cite{EMFF:JLQCD:Nf3:RG+Ovr}.
}
\label{tbl:chiral_fit:input:Ci}
\begin{tabular}{lllllll}
   $\Cppt\!\times\!10^5$  & $\Cpkt\!\times\!10^5$  & 
   $\Ctt\!\times\!10^5$  
   \\ \hline
   $-1.95(84)\left(^{+38}_{-21}\right)$   & 
   $-1.4(1.2)\left(^{+0.1}_{-0.7}\right)$  &
   $-6.4(1.1)(0.1)$                   
   \\ \hline
\end{tabular}
\end{center}
\vspace{0mm}
\end{table}
\end{ruledtabular}


The LO contribution to the other form factors are given as 
$\fmlo\!=\!0$ and $\fzlo\!=\!1$.
At higher orders, however, 
additional LECs appear through $f_-$,
which is absent in the EM form factors. 
For instance, the coefficients
\bea
   c_{-,t}^r
   & = &
   -2 C_{12}^r + C_{88}^r - C_{90}^r,
   \\
   c_{-,\pi}^r
   & = &
   6 C_{12}^r + 4 C_{13}^r + 2 C_{15}^r + 4 C_{17}^r + 2 C_{34}^r 
 +   C_{64}^r +   C_{65}^r +   C_{90}^r,
   \\
   c_{-,K}^r
   & = &
   6 C_{12}^r + 8 C_{13}^r + 4 C_{14}^r + 4 C_{15}^r + 2 C_{34}^r 
 + 2 C_{63}^r + 2 C_{64}^r +   C_{90}^r
\eea  
for the NNLO analytic terms for $\fm$ 
\bea
   \fmnnloC
   & = &
   4 c_{-,t}^r (M_K^2-M_\pi^2) t 
  +4 c_{-,\pi}^r (M_K^2-M_\pi^2) M_\pi^2
  +4 c_{-,K}^r (M_K^2-M_\pi^2) M_K^2
\eea
have $C_{14}^r$, $C_{15}^r$ and $C_{17}^r$.
The information of $\fp$ and the EM form factors is not so helpful 
in constraining them.
In this study, therefore,
we calculate the following quantity 
\bea
   \fzt(t)
   & = & 
   \fz(t) + \frac{t}{M_K^2-M_\pi^2}\left(1-\frac{F_K}{F_\pi}\right),
\eea
using our data of $F_K/F_\pi$ obtained in Ref.~\cite{Spectrum:Nf3:RG+Ovr:JLQCD}.
As proposed in Ref.~\cite{KFF:weak:ChPT:SU3:NNLO:BT},
the Callan-Treiman and Dashen-Weinstein theorems~\cite{Callan-Treiman,Dashen-Weinstein}
\bea
   \fz(M_K^2-M_\pi^2)
   & \sim &
   \frac{F_K}{F_\pi}
\eea
suggest 
a large cancellation between $\fz$ and $F_K/F_\pi$ 
even out of the Callan-Treiman point $t\!=\!M_K^2-M_\pi^2$.
Actually, in the chiral expansion of $\fzt$
\bea
   \fzt(t)
   & = &
   \fztlo + \fztnlo(t) + \fztnnlo(t) + \fztnnnlo(t),
   \label{eqn:chiral_fit:fzt}
   \\
   \fztnlo(t) 
   & = &
   \fztnloL(t) + \fztnloB(t),
   \\
   \fztnnlo(t) 
   & = &
   \fztnnloC(t) + \fztnnloL(t) + \fztnnloB(t),
\eea
the $L_i^r$-dependent NLO term vanishes, $\fztnloL\!=\!0$.
The NNLO analytic term has a rather simple form
\bea
   F_\pi^4\,\fztnnloC(t)
   & = &
  -8 (C_{12}^2+C_{34}^r) (M_K^2-M_\pi^2)^2 + 8 (2C_{12}^r + C_{34}^r) (M_K^2+M_\pi^2)t
  -8 C_{12}^r t^2
   \hspace{5mm}
   \nn \\
   & = &
  -8 \Cplspk (M_K^2-M_\pi^2)^2 + 8 (C_{12}^r + \Cplspk) (M_K^2+M_\pi^2)t
  -8 C_{12}^r t^2.
   \label{eqn:chiral_fit:fzt:nnlo_c}
\eea   
Genuine loop contributions, $\fztnloB$ and $\fztnnloB$, are free from 
the LECs, and we use the input in Table~\ref{tbl:chiral_fit:input:Li}
for $L_i^r$-dependent NNLO contribution $\fztnnloL$.
Therefore, a simultaneous fit to $\fp$ and $\fzt$ 
has only two fit parameters $\Cplspk$ and $C_{12}^r$ (or $C_{34}^r$).


\subsection{Chiral extrapolation and normalization of form factors}


\begin{figure}[t]
\begin{center}
   \includegraphics[angle=0,width=0.48\linewidth,clip]%
                   {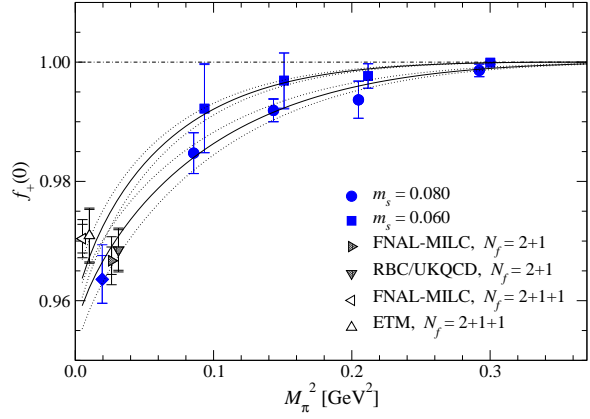}

   \vspace{0mm}
   \caption{
      Chiral extrapolation of $\fp(0)$ as a function of $M_\pi^2$. 
      Solid circles and squares show our data at $m_s\!=\!0.080$ and 0.060,
      respectively, whereas the solid diamond represents the value 
      extrapolated to the physical point $(m_{l,\rm phys},m_{s,\rm phys})$.
      We note that the physical strange quark mass $m_{s,\rm phys}$ (diamond)
      is slightly off the simulated values (solid lines).
      We also plot recent results for 
      $N_f\!=\!2+1$~\cite{Kl3:Nf3:FNAL+MILC,Kl3:Nf3:RBC/UKQCD} (shaded symbols) 
      and $N_f\!=\!2+1+1$~\cite{Kl3:Nf4:FNAL+MILC,Kl3:Nf4:ETM} (open symbols).
   }
   \label{fig:chiral_fit:f+0:nnlo}
\end{center}
\vspace{0mm}
\end{figure}

The Ademollo-Gatto theorem~\cite{Behrends-Sirlin,Ademollo-Gatto} states that
SU(3) breaking effects in $\fp(t)$
is second order in $(m_s\!-\!m_l)$ at $t\!=\!0$.
The chiral expansion~(\ref{eqn:chiral_fit:ff})\,--\,(\ref{eqn:chiral_fit:f+:Ctt}) is reduced into a simpler form with 
\bea
   \fpnloL(0)
   & = &
   0,
   \label{eqn:chiral_fit:f+0:fpnloL}
   \\
   \fpnloB(0)
   & = & 
   \frac{3}{2} H_{K \pi} + \frac{3}{2} H_{K \eta},
   \label{eqn:chiral_fit:f+0:fpnloB}
   \\
   \fpnnloC(0)
   & = & 
  -8\Cplspk (M_K^2-M_\pi^2)^2,
   \label{eqn:chiral_fit:f+0:fpnnloC}
\eea
where 
\bea
   H_{PQ}
   & = &
   - \frac{1}{128\pi^2F_{\pi}^2}
   \left\{ M_P^2 + M_Q^2 
         -\frac{2 M_P^2 M_Q^2}{M_P^2-M_Q^2} 
          \ln \left[ \frac{M_P^2}{M_Q^2} \right]
   \right\}.
   \label{eqn:chiral_fit:f+0:HPQ}
\eea
In previous lattice studies, therefore, 
one often determines $\fp(0)$ at simulated quark masses 
by assuming a phenomenological parametrization of 
the $t$ dependence of the form factors,
and then extrapolates $\fp(0)$ to the physical point 
$(m_{l,\rm phys},m_{s,\rm phys})$ based on NLO or NNLO ChPT.


\begin{ruledtabular}
\begin{table}[t]
\begin{center}
\caption{
   Numerical results of NNLO ChPT fits to form factors.
   The first line shows results of the conventional fit to $\fp(0)$
   in terms of the NG boson masses $M_{\{\pi,K\}}^2$.
   The second (third) line is from the fit to $\fp(t)$ ($\fp(t)$ and $\fzt(t)$)
   in terms of $M_{\{\pi,K\}}^2$ and $t$.
   The first error is statistical. 
   The second and third are systematics due to 
   the choice of the input $L_i^r$, 
   and truncation of the chiral expansion at NNLO.
   Note $\Cplspk\!=\!C_{12}^r+C_{34}^r$ (Eq.(\ref{eqn:chiral_fit:f+:Cplspk})).
}
\label{tbl:chiral_fit:result}
\begin{tabular}{l|llllll}
   fit data
     & $\fp(0)$  
     & $\Cplspk\!\times\!10^5$ 
     & $C_{12}^r\!\times\!10^5$
     & $C_{34}^r\!\times\!10^5$
   \\ \hline 
   $\fp(0)$ 
     & $0.9636(40)\left(^{+42}_{-6}\right)(+79)$ 
     & $0.53(7)(-6)(-17)$
     & --
     & --
   \\ \hline
   $\fp(t)$
     & $0.9691(42)\left(^{+45}_{-5}\right)(-45)$
     & $0.429(73)\left(^{+5}_{-71}\right)(+90)$
     & -- 
     & --
   \\ \hline
   $\fp(t)$, $\fzt(t)$
     & $0.9636(36)\left(^{+41}_{-4}\right)\left(^{+28}_{-19}\right)$
     & $0.524(62)\left(^{+1}_{-80}\right)\left(^{+33}_{-58}\right)$
     & $-0.23(7)\left(^{+34}_{-13}\right)\left(^{+5}_{-93}\right)$
     & $0.76(11)\left(^{+9}_{-42}\right)\left(^{+95}_{-11}\right)$
   \\ \hline 
\end{tabular}
\end{center}
\vspace{5mm}
\end{table}
\end{ruledtabular}


We carry out this type of the conventional analysis
using the data of $\fp(0)$ in Table~\ref{tbl:ff:q2_interp}.
As plotted in Fig.~\ref{fig:chiral_fit:f+0:nnlo},
the data are well described by the NNLO ChPT formula
with a good value of $\chi^2/{\rm d.o.f.}\!\sim\!0.2$.
Numerical fit results are summarized in Table~\ref{tbl:chiral_fit:result}.
We observe good agreement in $\fp(0)$ at the physical point
$(m_{l,\rm phys},m_{s,\rm phys})$ with recent lattice studies~\cite{Kl3:Nf3:FNAL+MILC,Kl3:Nf4:FNAL+MILC,Kl3:Nf3:RBC/UKQCD,Kl3:Nf4:ETM}.

\begin{figure}[t]
\begin{center}
   \includegraphics[angle=0,width=0.48\linewidth,clip]%
                   {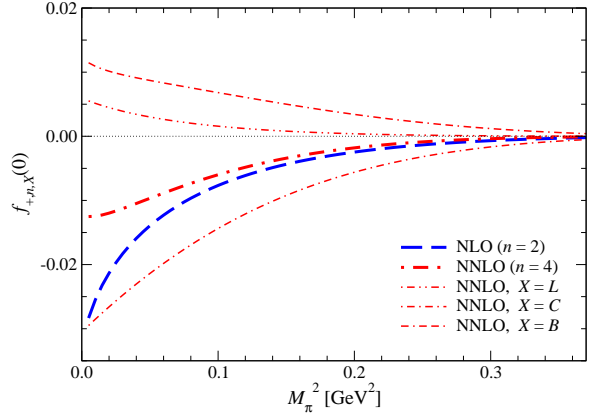}

   \vspace{0mm}
   \caption{
      LEC-(in)dependent NLO and NNLO contributions to $\fp(0)$.
      Thick dashed and dot-dashed lines are the NLO and NNLO contributions,
      whereas the NNLO terms, $\fpnnloL(0)$, $\fpnnloC(0)$ and $\fpnnloB(0)$,
      are plotted by the thin dot-dot-dashed, dot-dashed and 
      dot-dashed-dashed lines, respectively.
   }
   \label{fig:chiral_fit:f+0:contribu}
\end{center}
\vspace{0mm}
\end{figure}

The chiral expansion has reasonable convergence 
$\fp(0) \!=\! 0.9636(40) \!=\! 1 - 0.0232 -0.0132(40)$ at the physical point. 
However, Fig.~\ref{fig:chiral_fit:f+0:contribu} shows that 
the NLO and NNLO contributions, $\fpnlo(0)$ and $\fpnnlo(0)$,
are comparable at unphysically heavy $M_\pi\!\sim\!300$\,--\,500~MeV,
and there is a significant cancellation between 
the analytic ($\fpnnloC(0)$) 
and non-analytic ($\fpnnloL(0)$ and $\fpnnloB(0)$) NNLO contributions.

\begin{figure}[t]
\begin{center}
   \includegraphics[angle=0,width=0.48\linewidth,clip]%
                   {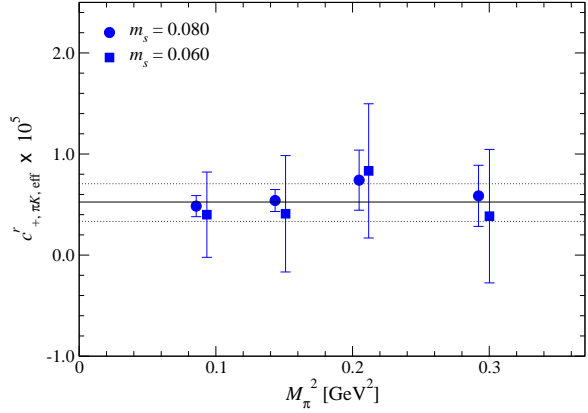}

   \vspace{0mm}
   \caption{
      Effective value $\Cplspkeff$ as a function of $M_\pi^2$. 
      Squares are slightly shifted along the horizontal axis for clarity.
      The solid and dotted lines show 
      $\Cplspk$ obtained from the NNLO chiral fit and its uncertainty.
   }
   \label{fig:chiral_fit:cpls_eff}
\end{center}
\vspace{0mm}
\end{figure}


In order to estimate the systematic error 
due to neglected higher order corrections,
we also test a fitting form including 
a N$^3$LO analytic term 
$\fpnnnlo\!=\!\Dpls\,M_\pi^2\,(M_K^2-M_\pi^2)^2 / F_\pi^6$,
where the factor $(M_K^2-M_\pi^2)^2$ is motivated by the Ademollo-Gatto theorem.
However, the coefficient $\Dpls\!=\!-1.1(1.6)\!\times\!10^6$ 
is poorly determined,
and the extrapolated value $\fp(0)\!=\!0.971(13)$ 
is statistically consistent with that from the NNLO fit.
This observation and the good value of $\chi^2/{\rm d.o.f.}$ for the NNLO fit 
suggest that 
the uncertainty due to the truncation of the chiral expansion at NNLO 
is not large compared to the statistical accuracy.
We treat the difference in $\fp(0)$ 
between the fits with and without the N$^3$LO term  
as a systematic uncertainty in Table~\ref{tbl:chiral_fit:result}.


We can examine the significance of the higher order correction 
without assuming the form of $\fpnnnlo$. Let us consider a quantity
\bea
   \Delta f(0)
   & = & 
   \fp(0) - \fplo - \fpnloL(0) - \fpnloB(0) - \fpnnloL(0) - \fpnnloB(0),
\eea 
which is the sum of the NNLO analytic term and 
the possible higher order correction $\fpnnloC(0)\!+\!\fpnnnlo(0)$.
Note that 
$\fplo(0)$, $\fpnloB(0)$ and $\fpnnloB(0)$ are LEC-free in $\xi$-expansion,
and hence 
$\Delta f$ can be calculated from our data of $\fp(0)$ and 
inputs in Tables~\ref{tbl:chiral_fit:input:Li}.
We can define an effective value of $\Cplspk\!=\!C_{12}^r+C_{34}^r$ as 
\bea
   \Cplspkeff
   & = & 
   -\frac{F_\pi^4}{8(M_K^2-M_\pi^2)^2} \Delta f
   = \Cplspk + O(M_\pi^2,M_K^2),
\eea
which deviates from $\Cplspk$ and shows a non-trivial quark mass dependence,
if the higher order correction $\fpnnnlo$ is significant 
in the simulation region.
As shown in Fig.~\ref{fig:chiral_fit:cpls_eff}, however, 
our result has small dependence on $m_l$ and $m_s$ 
suggesting that 
the higher order correction is not large 
compared to the statistical accuracy.


\begin{figure}[t]
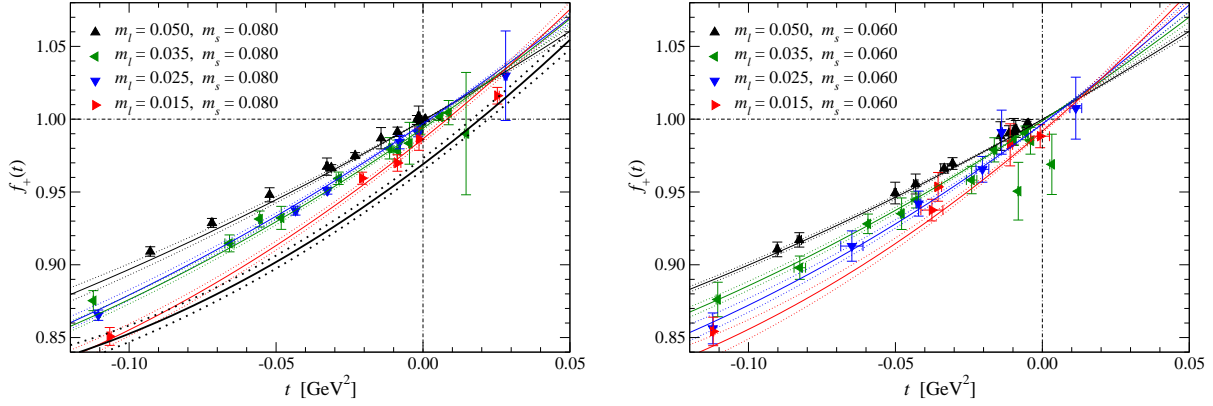

\begin{center}
   \includegraphics[angle=0,width=0.48\linewidth,clip]%
                   {f+_vs_q2.ms0080.phys.su3.w_Fpi.nnlo.be14.eps}
   \hspace{3mm}
   \includegraphics[angle=0,width=0.48\linewidth,clip]%
                   {f+_vs_q2.ms0060.phys.su3.w_Fpi.nnlo.be14.eps}

   \vspace{0mm}
   \caption{
      Chiral extrapolation of $\fp(t)$ as a function of $t$.
      Different symbols show data at different values of $m_l$,
      whereas the left and right panels are for $m_s\!=\!0.080$ and 0.060,
      respectively.
      Thin solid and dotted lines show the fit curve and its statistical 
      uncertainty. We also plot those at the physical point 
      $(m_{l,\rm phys},m_{s,\rm phys})$ by thick lines 
      in the left panel at $m_s\!\sim\!m_{s, \rm phys}$.
   }
   \label{fig:chiral_fit:f+:nnlo}
\end{center}
\vspace{0mm}
\end{figure}

It is advantageous to use not only $\fp(0)$ 
but all the data of $\fp(t)$ 
to better constrain the possible higher order chiral corrections. 
Parametrizing both the $t$ and quark mass dependences based on ChPT 
reduces the model dependence of our analysis.
We therefore carry out a fit for $f_+(t)$ 
using the chiral expansion (\ref{eqn:chiral_fit:ff})
as a function of $t$, $M_\pi^2$ and $M_K^2$. 
As shown in Fig.~\ref{fig:chiral_fit:f+:nnlo},
the $t$ dependence of our data is also described well by the NNLO formula
with an acceptable value of $\chi/{\rm d.o.f.}\!\sim\!0.7$, 
because we simulate a limited region of $t\!\sim\!0$.


Results for the LEC $\Cplspk$ and the normalization $\fp(0)$ 
at the physical point in Table~\ref{tbl:chiral_fit:result} 
show good consistency with those from the conventional analysis.
Their systematic error due to the 
truncation of the chiral expansion at NNLO
is estimated by repeating the fit 
with each of the following higher order terms  
\bea
   F_\pi^6\, \fpnnnlo
   & = & 
   \Dpls M_\pi^2 (M_K^2-M_\pi^2)^2,
   \hspace{3mm}
   \Dpls M_\pi^4 t, 
   \hspace{3mm}
   \Dpls M_\pi^2 t^2, 
   \hspace{3mm}
   \Dpls M_\pi^2 M_K^2 t.
   \label{eqn:chiral_fit:f+:n3lo}
\eea
This uncertainty 
is slightly smaller than the conventional analysis, 
because the coefficient $\Dpls$ is better constrained 
with more data at non-zero $t$'s.

\begin{figure}[t]
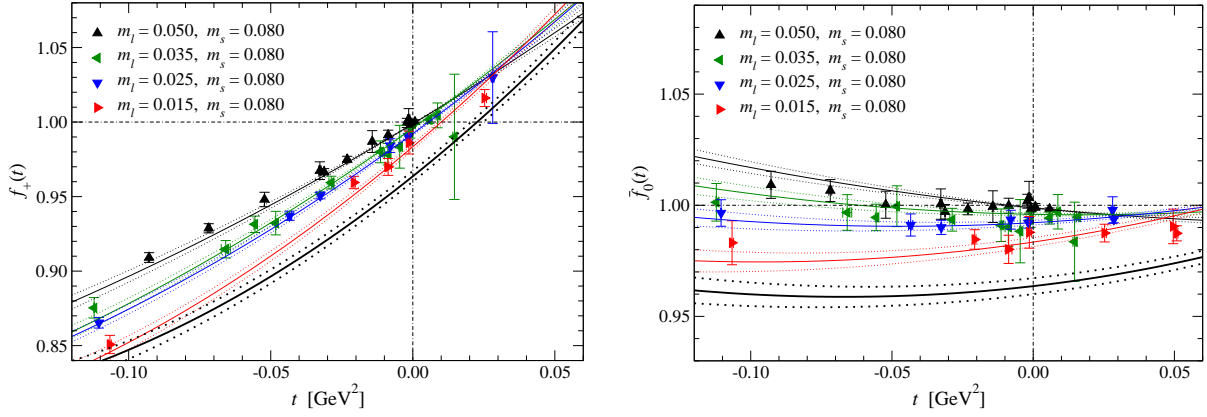

\begin{center}
   \includegraphics[angle=0,width=0.48\linewidth,clip]%
                   {f+_vs_q2.ms0080.f+_f0t.phys.su3.w_Fpi.nnlo.be14.eps}
   \hspace{3mm}
   \includegraphics[angle=0,width=0.48\linewidth,clip]%
                   {f0t_vs_q2.ms0080.f+_f0t.phys.su3.w_Fpi.nnlo.be14.eps}

   \vspace{0mm}
   \caption{
      Simultaneous chiral fit to $\fp(t)$ (left panel) and 
      $\fzt(t)$ (right panel) at $m_s\!=\!0.080$.
      Data at different values of $m_l$ are plotted by different symbols 
      as a function of $t$.
      Thin solid and dotted line shows the fit curve and the statistical 
      uncertainty, whereas thick lines show those 
      at $(m_{l,\rm phys},m_{s,\rm phys})$. 
   }
   \label{fig:chiral_fit:f+_f0t:nnlo}
\end{center}
\vspace{0mm}
\end{figure}


In order to make use of all the available data,
we performed a simultaneous fit to $\fp(t)$ and $\fzt(t)$ 
as a function of $t$, $M_\pi^2$ and $M_K^2$. 
As shown in Fig.~\ref{fig:chiral_fit:f+_f0t:nnlo},
the NNLO formulae (\ref{eqn:chiral_fit:ff}) and (\ref{eqn:chiral_fit:fzt})
describe our data well 
with a good value of $\chi^2/{\rm d.o.f.}\!\sim\!0.7$.
Numerical results of the fit are summarized 
in Table~\ref{tbl:chiral_fit:result}.
We estimate the uncertainty due to possible higher order corrections 
by testing the N$^3$LO terms (\ref{eqn:chiral_fit:f+:n3lo}) for $\fpnnnlo$
and the followings for $\fztnnnlo$
\bea
   F_\pi^6\, \fztnnnlo
   & = & 
   \Dpls M_\pi^2 (M_K^2-M_\pi^2)^2,
   \hspace{3mm}
   \Dz M_\pi^4 t, 
   \hspace{3mm}
   \Dz M_\pi^2 t^2, 
   \hspace{3mm}
   \Dz M_\pi^2 M_K^2 t.
   \label{eqn:chiral_fit:f+_f0t:n3lo}
\eea

\begin{figure}[t]
\begin{center}
   \includegraphics[angle=0,width=0.48\linewidth,clip]%
                   {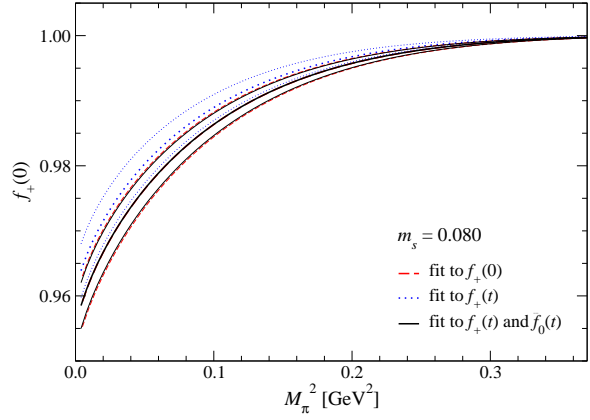}
   \vspace{0mm}
   \caption{
      Comparison of fit curves for $\fp(0)$ 
      at $m_s\!=\!0.080\!\sim\!m_{s,\rm phys}$. 
      The dashed, dotted and solid lines lines are 
      reproduced from the fits to $\fp(0)$, $\fp(t)$
      and the simultaneous fit to $\fp(t)$ and $\fzt(t)$, respectively.
      For each fit,
      the thick line shows the central value, 
      whereas its statistical error is shown by the two thin lines.
   }
   \label{fig:chiral_fit:f+0:comparison}
\end{center}
\vspace{0mm}
\end{figure}

Table~\ref{tbl:chiral_fit:result} shows good consistency 
in $\Cplspk$ and $\fp(0)$ at the physical point 
among the three types of the chiral fit: 
namely, the fit to $\fp(0)$, that to $\fp(t)$,
and the simultaneous fit to $\fp(t)$ and $\fzt(t)$.
This is also demonstrated in Fig.~\ref{fig:chiral_fit:f+0:comparison},
where the $M_\pi$ dependence of $f_+(0)$ at $m_s\!=\!0.080\!\sim\!m_{s,\rm phys}$
is reproduced from the three fits.
We observe good agreement within $\sim\!1~\sigma$
in the whole simulation region of $M_\pi^2$.

Table~\ref{tbl:chiral_fit:result} shows that 
the statistical accuracy of the fit results is not largely different
between the conventional fit and the fit to $\fp$.
Indeed, these two fits use the same data of $\fp$
but different parametrizations for the $t$ dependence:
Eq.~(\ref{eqn:ff:t-dep:vmd+poly}) or ChPT.
The statistical error is slightly reduced by using $\fzt$. 
The uncertainty due to the choice of the input 
(Tables~\ref{tbl:chiral_fit:input:Li} and \ref{tbl:chiral_fit:input:Ci})
is more or less the same among the three fits. 
However, 
the uncertainty due to the truncation of the chiral expansion is
significantly reduced by including more data into the ChPT fit.
From this observation,
we consider the simultaneous fit to $\fp$ and $\fzt$ as our best fit.


\begin{figure}[t]
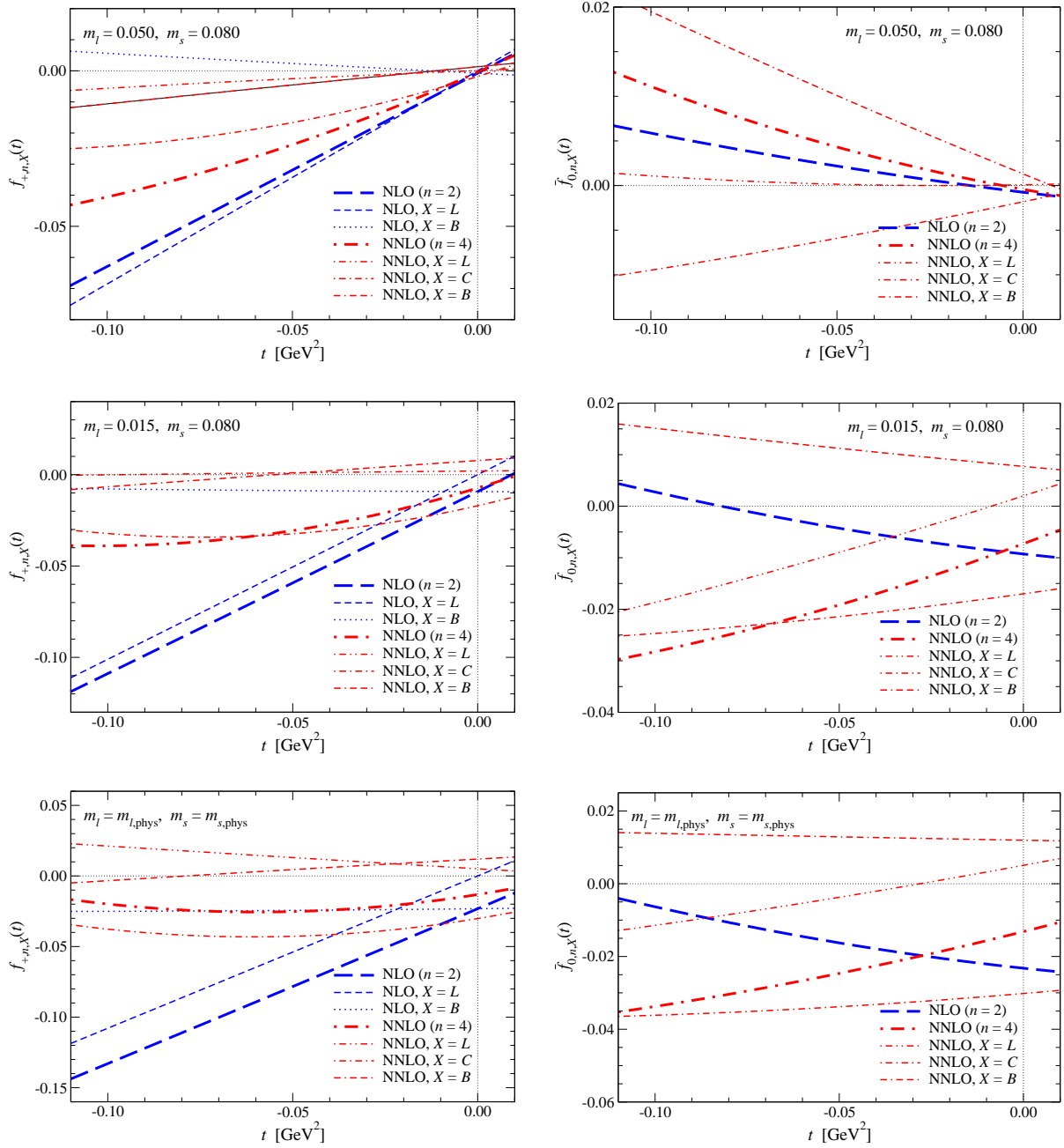

\begin{center}
   \includegraphics[angle=0,width=0.48\linewidth,clip]%
                   {f+_vs_q2_mud0050_ms0080.phys.su3.w_Fpi.nnlo.be14.cntrb.eps}
   \hspace{3mm}
   \includegraphics[angle=0,width=0.48\linewidth,clip]%
                   {f0t_vs_q2_mud0050_ms0080.phys.su3.w_Fpi.nnlo.be14.cntrb.eps}

   \vspace{5mm}
   \includegraphics[angle=0,width=0.48\linewidth,clip]%
                   {f+_vs_q2_mud0015_ms0080.phys.su3.w_Fpi.nnlo.be14.cntrb.eps}
   \hspace{3mm}
   \includegraphics[angle=0,width=0.48\linewidth,clip]%
                   {f0t_vs_q2_mud0015_ms0080.phys.su3.w_Fpi.nnlo.be14.cntrb.eps}

   \vspace{5mm}
   \includegraphics[angle=0,width=0.48\linewidth,clip]%
                   {f+_vs_q2_mudphys_msphys.phys.su3.w_Fpi.nnlo.be14.cntrb.eps}
   \hspace{3mm}
   \includegraphics[angle=0,width=0.48\linewidth,clip]%
                   {f0t_vs_q2_mudphys_msphys.phys.su3.w_Fpi.nnlo.be14.cntrb.eps}

   \vspace{0mm}
   \caption{
      LEC-(in)dependent NLO and NNLO contributions
      to $\fp(t)$ (left panels) and $\fzt(t)$ (right panels)
      as a function of $t$.
      The top, middle and bottom panels show data at 
      $(m_l,m_s)\!=\!(0.050,0.080)$, (0.015,0.080)
      and the physical point $(m_{l,\rm phys},m_{s,\rm phys})$,
      respectively.
      The net NLO and NNLO contributions are plotted by 
      thick blue dashed and red dot-dashed lines, respectively.
      Thin lines show their breakdown into 
      LEC-(in)dependent terms.
   }
   \label{fig:chiral_fit:f+_f0t:contribu}
\end{center}
\vspace{0mm}
\end{figure}


We investigate the convergence of this best fit
in Fig.~\ref{fig:chiral_fit:f+_f0t:contribu}.
Our observations on $\fp$ at 
large space-like momentum transfer  $-t\!\gg\!0$
are similar to those on the charged meson EM form factors $\pkpff$
in Ref.~\cite{EMFF:JLQCD:Nf3:RG+Ovr}.
The analytic term $\fpnloL$ ($\fpnnloC$) is dominant NLO 
(largest NNLO) contribution to $\fpnlo$ ($\fpnnlo$).
We note that 
our estimate of the relevant LECs 
in Tables~\ref{tbl:chiral_fit:input:Li}\,--\,\ref{tbl:chiral_fit:result} 
is not unexpectedly large,
and consistent with an order estimate~\cite{ChPT:LECs:SU2+SU3}
\bea
   L_i^r = O((4\pi)^{-2}) = O(6\!\times\!10^{-3}),
   \hspace{5mm}
   C_i^r = O((4\pi)^{-4}) = O(4\!\times\!10^{-5}).
   \label{eqn:chiral_fit:LEC_order}
\eea


The large analytic terms, $\fpnloL$ and $\fpnnloC$, 
suggest the importance of the first-principle determination of 
the relevant LECs. 
While we employ the input in Table~\ref{tbl:chiral_fit:input:Li}
for other LECs involved in $\fpnnloL$,
this term turns out to be small.
Therefore, the systematic uncertainty 
due to the choice of the input is not large.


The convergence at $-t\!\gg\!0$ 
is reasonably good and becomes better toward smaller $M_\pi$.
This is because the dominant NLO term $\fpnloL\!\propto\!t/F_\pi^2$ 
is not suppressed by the NG boson masses and 
even enhanced by the factor $F_\pi^{-2}$ in the $\xi$-expansion.
A similar convergence property is also observed 
for the EM form factors $\pkpff$~\cite{EMFF:JLQCD:Nf3:RG+Ovr}.

There is a property of $\fp$ different from $\pkpff$ 
near the important reference point $t\!=\!0$.
Unless $m_l\!\ne\!m_s$,
each chiral correction does not necessarily vanish except $\fpnloL$,
which is dominantly large at $-t\!\gg\!0$
(blue thin dashed lines in Fig.~\ref{fig:chiral_fit:f+_f0t:contribu}).
As a result,
the chiral expansion has a poorer convergence towards $t\!=\!0$
as already observed in Fig.~\ref{fig:chiral_fit:f+0:contribu}.
Note, however, that our analysis in Fig.~\ref{fig:chiral_fit:cpls_eff}
does not suggest statistically significant N$^3$LO nor higher order 
corrections.


%
%


%
%

%
%


In our analysis,
the quantity $\fzt$ is used to obtain additional constraints 
on the relevant $O(p^6)$ couplings, $C_{12}^r$ and $C_{34}^r$. 
For this purpose, 
$\fzt$ is designed to have no NLO analytic term, 
which is a dominant contribution to $\fp$,
and its dependence on $L_i^r$ starts from NNLO.
The right panels of Fig.~\ref{fig:chiral_fit:f+_f0t:contribu} show
that the $L_i$-dependent NNLO correction $\fztnnloL$ is not large
with our choice of the input in Table~\ref{tbl:chiral_fit:input:Li}.
The remaining loop corrections, $\fztnloB$ and $\fztnnloB$, 
are parameter-free in the $\xi$-expansion.
Therefore, 
$\fzt$ has reasonable sensitivity to the NNLO analytic term $\fztnnloC$.
This leads to our observation in Table~\ref{tbl:chiral_fit:result}:
incorporating $\fzt$ into the chiral extrapolation 
is helpful in improving the statistical accuracy of $\Cplspk$ and $\fp(0)$,
and also in reducing their systematic error 
due to the truncation of the chiral expansion.



For phenomenological applications, 
it is preferable to separately fix $C_{12}^r$ and $C_{34}^r$
rather than their sum $\Cplspk$.
This is only possible with the data of $\fzt$
by disentangle their $M_\pi^2$, $M_K^2$ and $t$ dependences 
in the NNLO correction (\ref{eqn:chiral_fit:fzt:nnlo_c}).
As mentioned above, however, 
$\fzt$ is designed to have the small NLO correction $\fztnlo$. 
This quantity is sensitive to not only NNLO but 
the even higher order corrections 
(\ref{eqn:chiral_fit:f+:n3lo}) and (\ref{eqn:chiral_fit:f+_f0t:n3lo})
added by hand to estimate the systematic uncertainty
due to the truncation of the chiral expansion.
As a result,
this uncertainty for $C_{12}^r$ and $C_{34}^r$ 
is rather large in Table~\ref{tbl:chiral_fit:result}.
In this study, therefore,
we only confirm that our results are 
consistent with the order estimate~(\ref{eqn:chiral_fit:LEC_order}).
We note that a better determination of $C_{12}^r$ and $C_{34}^r$
needs more data of $\fzt$ at smaller values of $M_\pi$ and $M_K$.

%
%
%
%
%
%


\subsection{Form factor shape}


In order to support the reliability of the determination of $\fp(0)$
with the sub-percent level accuracy,
it is important to check the consistency of the form factors' shape 
on the lattice with experiments. 
In recent analyses of experimental data,
it is popular to employ the so-called 
dispersive parametrization of the $t$ dependence~\cite{shape:disp:scalar,shape:disp:vector} 
based on the analyticity of the form factors.
In this study, however, we consider the slope $\lambdapzp$ 
in the conventional quadratic parametrization~(\ref{eqn:ff:t-dep:poly}).
These are convenient in our analysis based on ChPT,
since the chiral expansion is the expansion in terms of $t$
(and the NG boson masses).
We note that the quadratic parametrization has been also 
well studied phenomenologically and experimentally.
Its relation to the dispersive one has been established~\cite{shape:disp:vector}.
From recent experimental data~\cite{CKM16:Moulson},
the slopes are estimated as 
\bea
   &&
   \lambdapp = 2.58(7) \times 10^{-2},
   \hspace{5mm}
   \lambdazp = 1.36(7) \times 10^{-2}.
   \label{eqn:chiral_fit:slope:exp}
\eea

In this paper, we treat $M_{\pi^\pm}^2$ in Eq.~(\ref{eqn:ff:t-dep:poly})
just as the normalization factor 
to make $\lambda_{\{+,0\}}$ dimensionless,
and fix it to its physical value.
The slope is then given as 
\bea
   \lambdapzp
   & = &
   \left.
      \frac{M_{\pi^\pm,\rm phys}^2}{\fpz(0)} \frac{d \fpz(t)}{dt}
   \right|_{t=0}.
   \label{eqn:chiral_fit:slope:slope}
\eea
We evaluate 
both the normalization $\fp(0)\!=\!\fz(0)$ (Table~\ref{tbl:chiral_fit:result})
and derivatives $df_{\{+,0\}}/dt|_{t=0}$
from our chiral fit of $\fp$ and $\fzt$ based on NNLO ChPT.


It is straightforward to calculate $d\fp/dt|_{t=0}$.
From the chiral expansion (\ref{eqn:chiral_fit:ff}),
(\ref{eqn:chiral_fit:fp:nlo}) and (\ref{eqn:chiral_fit:fp:nnlo}),
it is given as 
\bea
   \dfdtp
   & = &
   \dfdtpnlo + \dfdtpnnlo + \dfdtpnnnlo,
   \\
   \dfdtpnlo
   & = & 
   \dfdtpnloL + \dfdtpnloB,
   \\
   \dfdtpnnlo
   & = & 
   \dfdtpnnloC +\dfdtpnnloL + \dfdtpnnloB.
\eea
The derivatives of $\fpnloL$, $\fpnloB$, $\fpnnloC$ and $\fpnnnlo$ 
are analytically calculable from their expressions
(\ref{eqn:chiral_fit:fp:nlo_l}), (\ref{eqn:chiral_fit:fp:nlo_b}), 
(\ref{eqn:chiral_fit:fp:nnlo_c}) and (\ref{eqn:chiral_fit:f+:n3lo}).
Since the NNLO non-analytic terms, $\fpnnloL$ and $\fpnnloB$,
have rather lengthy expressions,
we numerically evaluate their derivatives
as in our study of the light meson charge radii~\cite{EMFF:JLQCD:Nf3:RG+Ovr}.


We evaluate $d\fz/dt|_{t=0}$ through
\bea
   \dfdtz
   & = &
   \dfdtzt
 - \frac{1}{M_K^2-M_\pi^2}
   \left( 1 - \frac{F_K}{F_\pi} \right).
   \label{eqn:chiral_fit:slope:df0dt}
\eea   
Here $d\fzt/dt|_{t=0}$ is calculated in a similar way to $d\fp/dt|_{t=0}$,
whereas $F_K/F_\pi$ is estimated from 
our NNLO chiral fit in Ref.~\cite{Spectrum:Nf3:RG+Ovr:JLQCD}.


\begin{figure}[t]
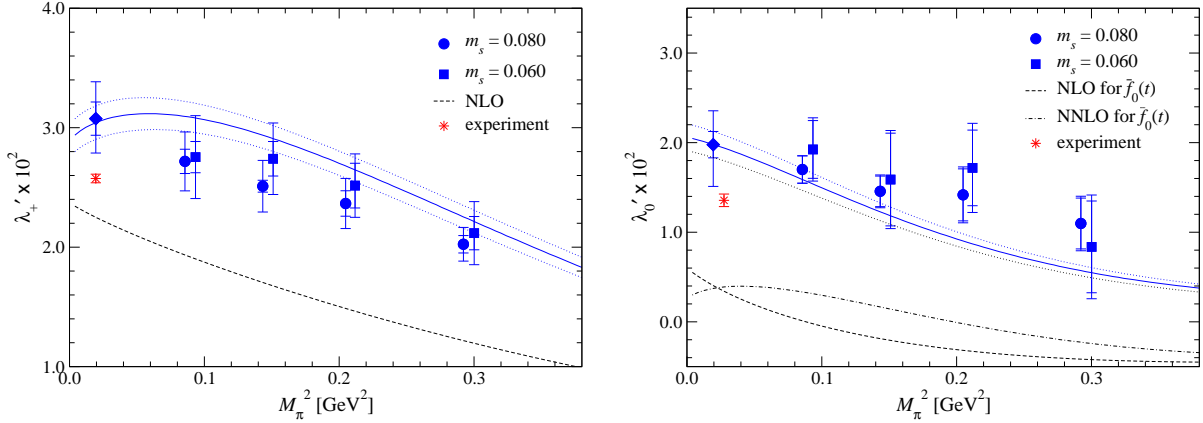

\begin{center}
   \includegraphics[angle=0,width=0.48\linewidth,clip]%
                   {lambda+p_vs_Mpi2.eps}
   \hspace{3mm}
   \includegraphics[angle=0,width=0.48\linewidth,clip]%
                   {lambda0p_vs_Mpi2.eps}
   \vspace{0mm}
   \caption{
     Slope parameters $\lambdapp$ (left panel) 
     and $\lambdazp$ (right panel) as a function of $M_\pi^2$.
     The blue solid and dotted lines are reproduced from our chiral fit
     based on NNLO ChPT, and represent the slopes 
     at a simulated strange quark mass $m_s\!=\!0.080$.
     The black dashed (dot-dashed) line shows the contribution 
     from the NLO (NNLO) correction to $\fp$ and $\fzt$.
     The value extrapolated to the physical point 
     $(m_{ud,\rm phys},m_{s,\rm phys})$ is plotted by the blue diamonds,
     whereas the red stars represent 
     the experimental values~(\ref{eqn:chiral_fit:slope:exp}).
     We also plot the values in Table~\ref{tbl:ff:q2_interp} 
     by blue circles ($m_s\!=\!0.080$) and squares (0.060).
   }
   \label{fig:chiral_fit:slope}
\end{center}
\vspace{0mm}
\end{figure}

Our results for $\lambdapzp$ are plotted in Fig.~\ref{fig:chiral_fit:slope}
as a function of $M_\pi^2$.
We observe reasonable agreement with 
the values in Table~\ref{tbl:ff:q2_interp}, which are estimated by assuming 
the VMD-based parametrization~(\ref{eqn:ff:t-dep:vmd+poly}) for $f_+$
and the quadratic form (\ref{eqn:ff:t-dep:poly}) for $f_0$.
This agreement does not necessarily hold,
since the non-analytic chiral behavior is not explicitly taken into account
in the model assumptions.
The reasonable consistency in $\lambdapp$ is therefore 
compatible with our observation in Fig.~\ref{fig:chiral_fit:f+_f0t:contribu}
that non-analytic corrections to $f_+$ are not large at the simulation points.


The dashed line in the left panel of Fig.~\ref{fig:chiral_fit:slope}
shows $\lambdapp$ up to NLO.
The NNLO correction turns out to be significant 
at the simulation points and down to the physical point.
In the whole region,
the analytic term $\fpnnloC$ gives rise to a dominant part of the NNLO correction
suggesting the importance of the first-principle determination of 
the relevant LECs. 
We observe that
contribution from the NNLO non-analytic term $\fpnnloL$ become significant
below the simulation points $M_\pi\!\lesssim\!0.09~\mbox{GeV}^2$
and leads to the non-monotonous $M_\pi^2$ dependence of $\lambdapp$.
It is therefore important to study the form factor shape 
by taking account of the chiral logarithmic terms.


In the right panel of Fig.~\ref{fig:chiral_fit:slope},
dashed and dot-dashed lines show contributions to $\lambdazp$ from $\fzt$,
which are not large.
The slope is therefore dominated by the second term 
in the right-hand side of Eq.~(\ref{eqn:chiral_fit:slope:df0dt}).
This is because a large part of the NLO and NNLO analytic terms, 
which give rise to a large contribution to $\lambdapp$,
are absorbed into the second term. 
This suggests that a modified version~\cite{KFF:ChPT:SU3:NLO} 
of the Dashen-Weinstein relation~\cite{Dashen-Weinstein}
\bea
   \frac{\fz^\prime(0)}{\fz(0)}
   & = &
   \frac{\lambdazp}{M_{\pi^\pm,\rm phys}^2}  
   \sim\ \frac{1}{M_K^2-M_\pi^2}\left(\frac{F_K}{F_\pi}-1\right)
\eea
holds reasonably well in a wide region of $M_\pi\!\lesssim\!500$~MeV.


\subsection{Numerical results for form factors and LEC}


Our numerical result for the normalization is 
\bea
   \fp(0)
   & = &
   0.9636(36)_{\rm stat} \left(^{+49}_{-19}\right)_{\rm chiral} (29)_{a\ne 0}.
\eea
The first error is statistical. 
The second error is due to the chiral extrapolation
and is a quadrature sum of the uncertainties 
from the choice of the input $L_{\{1,...,8\}}$ 
(Table~\ref{tbl:chiral_fit:input:Li})
and the truncation of the chiral expansion at NNLO.
The third one represents discretization errors at our finite lattice spacing.
Since the LO term $\fplo\!=\!1$ is fixed from symmetry, 
we assign discretization errors to the non-trivial chiral correction
$\fpnlo+\fpnnlo$
by an order counting $O((a\Lambda_{\rm QCD})^2)\!\sim\!8$\,\%
with $\Lambda_{\rm QCD}\!=\!500$~MeV.
We expect that 
$\fpnlo+\fpnnlo$ also receives finite volume effects of 
$O(e^{-M_\pi L})\!\sim\!1$\,--\,2\,\%.
This is, however, well below other uncertainties.

A latest analysis of available experimental data 
together with analytic calculations of the isospin 
and EM corrections~\cite{CKM16:Moulson,Kl3:EM:CKNRT,Kl3:isospin:KN,Kl3:EM:CGN}
obtains $|V_{us}|\fp(0)\!=\!0.21654(41)$~\cite{CKM16:Moulson}.
Our results leads to 
\bea
   |V_{us}|
   & = &
   0.2247\left(^{+16}_{-12}\right)_{\rm th}(4)_{\rm ex},
\eea
and a measure of the unitarity violation in the first row
\bea
   \Delta_{\rm CKM}
   & = & 
   |V_{ud}|^2+|V_{us}|^2+|V_{ub}|^2-1
   = 
   -0.0004\left(^{+7}_{-8}\right).
\eea
Here we use recent estimate $|V_{ud}|\!=\!0.97420(21)$~\cite{CKM16:Hardy}
from the super-allowed nuclear $\beta$ decays.
Note that $|V_{ub}|\!\approx\!4\!\times\!10^{-3}$~\cite{PDG16}
is too small to affect this test of CKM unitarity,
and the long-standing tension between the exclusive and 
inclusive decays does not change $\Delta_{\rm CKM}$ significantly.
CKM unitarity fulfilled at the level of $O(0.1\,\%)$ may have sensitivity 
to new physics at the TeV scale~\cite{NPtest:GC}.


For the LEC and form factor shape, we obtain
\bea
   \Cplspk(M_\rho)
   & = & 
    C_{12}^r(M_\rho) + C_{34}^r(M_\rho) 
   =
   0.524(62)_{\rm stat}\left(^{+33}_{-99}\right)_{\rm chiral} (42)_{a\ne0}
   \times 10^{-5},
   \\
   \lambdapp
   & = & 
   3.08(14)_{\rm stat} \left(^{+12}_{-4}\right)_{\rm chiral} (25)_{a\ne0} 
   \times 10^{-2},
   \\   
   \lambdazp
   & = & 
   1.98(15)_{\rm stat} \left(^{+31}_{-41}\right)_{\rm chiral} (16)_{a\ne0} 
   \times 10^{-2},
\eea
where we assign $O((a\Lambda_{\rm QCD})^2)$ discretization errors,
and 1\,--\,2\,\% finite volume effects are well below other uncertainties.
We note that 
our chiral fit to $\fp(0)$ and that to $\fp(t)$ yield 
consistent results for the normalization and LEC.
Recent lattice estimate~\cite{Kl3:Nf3:FNAL+MILC,Kl3:Nf3:RBC/UKQCD,Kl3:Nf4:FNAL+MILC,Kl3:Nf4:ETM}
and the current world average~\cite{FLAG3} 
for $\fp(0)$ are in good agreement with our result.
We also observe reasonable consistency
with a previous lattice estimate of the LEC
$\Cplspk(M_\rho)\!=\!0.46(10)\!\times\!10^{-5}$~\cite{Kl3:Nf3:FNAL+MILC}
and experimental results for the slopes (\ref{eqn:chiral_fit:slope:exp}).


\section{Conclusions} 
\label{sec:conclusion}


In this article,
we have presented our study of the chiral behavior of the 
$K\!\to\!\pi$ semileptonic form factors.
Relevant meson correlators are precisely calculated 
by using the all-to-all quark propagator.
Our data of the form factors are directly compared 
with NNLO ChPT in the continuum limit 
by exploiting exact chiral symmetry preserved by the overlap quark action.


Similar to our observation in our study of the EM form factors,
the non-trivial chiral correction to the vector form factor 
$\fp\!-\!1$ is largely dominated 
by the NLO analytic term $\fpnloL\!\propto\!t$.
As a result, the NNLO chiral expansion exhibits 
reasonably good convergence particularly at $t\!\ll\!0$,
and describes our data reasonably well.
While $\fpnloL$ vanishes and the convergence becomes poorer towards $t\!=\!0$,
our analysis suggests that N$^3$LO and even higher order corrections
are not large compared to the statistical accuracy.


We determined the normalization $\fp(0)$ within $\leq\!1$\,\% accuracy
from our chiral fit to $\fp(t)$ and $\fzt(t)$ based on NNLO ChPT.
The result is nicely consistent with CKM unitarity in the first row.
We also estimate the relevant $O(p^6)$ coupling $C_{12}^r\!+\!C_{34}^r$ 
and the slope parameters $\lambdapzp$ from the same fit, 
and observe reasonable consistency with a previous lattice study
and experiments, respectively.


The statistical and systematic uncertainties of $\fp(0)$ 
turn out to be comparable to each other with our simulation set-up.
Its accuracy can be improved in the future 
by more realistic simulations with higher statistics. 
Note that
the uncertainty of the isospin breaking and EM corrections
are typically 0.1\,--\,0.2\%.
Good control of these corrections will be increasingly important,
as the accuracy of $\fp(0)$ approaches this level.


It is interesting to extend this work to 
the heavy meson decays to test the Standard Model
by a broad range of experimental measurements.
Simulations on finer lattices are underway~\cite{JLQCD:Noaki:Lat14}
by using a computationally inexpensive fermion formulation
with good chiral symmetry~\cite{JLQCD:TK:Lat13}.
Preliminary results have been reported 
for the $D$ and $B$ meson decay constants~\cite{Lat16:Fahy}
and the $D\!\to\!\pi$ and $D\!\to\!K$ form factors~\cite{Lat16:Kaneko}.


\begin{acknowledgments}

We thank Johan Bijnens for making his code 
to calculate the form factors in NNLO SU(3) ChPT
available to us.
Numerical simulations were performed on Hitachi SR16000 and 
IBM System Blue Gene Solution 
at High Energy Accelerator Research Organization (KEK) 
under a support of its Large Scale Simulation Program (No.~16/17-14),
and on SR16000 at YITP in Kyoto University.
This work is supported in part by JSPS KAKENHI Grant Numbers
JP25800147, JP26400259, JP16H03978
and by MEXT as ``Priority Issue on Post-K computer''
(Elucidation of the Fundamental Laws and Evolution of the Universe) and JICFuS.

\end{acknowledgments}



\begin{thebibliography}{99}
\bibitem{KFF:weak:ChPT:SU3:NNLO:BT:loop} 
The formulae for the loop corrections, 
$\fpnnloL$, $\fpnnloB$, $\fmnnloL$ and $\fmnnloB$,
are available from the authors of Ref.~\cite{KFF:weak:ChPT:SU3:NNLO:BT} 
on request.
\end{thebibliography}
\end{document}